\documentclass[num-refs]{nbdt-article}

\usepackage{siunitx}

\pdfoutput=1

\papertype{Original Article}
\paperfield{Behavior}

\title{ 
Visuomotor feedback tuning in the absence of visual error information}
\abbrevs{EMG, electromyography; CW, clockwise; CCW, counterclockwise; MPE, maximum perpendicular error; ANOVA, analysis of variance.}

\author[1]{Sae Franklin}
\author[1,2,3]{David W. Franklin}
\affil[1]{Neuromuscular Diagnostics, Department Health and Sport Sciences, TUM School of Medicine and Health, Technical University of Munich, Germany}
\affil[2]{Munich Institute of Robotics and Machine Intelligence (MIRMI), Technical University of Munich, Germany}
\affil[3]{Munich Data Science Institute (MDSI), Technical University of Munich, Munich, Germany}

\corraddress{David W. Franklin, Neuromuscular Diagnostics, Department Health and Sport Sciences, TUM School of Medicine and Health, Technical University of Munich, Munich, 80992, Germany}
\corremail{david.franklin@tum.de}

\runningauthor{Franklin and Franklin}

\begin{document}

\maketitle

\begin{abstract}
Large increases in visuomotor feedback gains occur during initial adaptation to novel dynamics, which we propose are due to increased internal model uncertainty. That is, large errors indicate increased uncertainty in our prediction of the environment, increasing feedback gains and co-contraction as a coping mechanism. Our previous work showed distinct patterns of visuomotor feedback gains during abrupt or gradual adaptation to a force field, suggesting two complementary processes: reactive feedback gains increasing with internal model uncertainty and the gradual learning of predictive feedback gains tuned to the environment. Here we further investigate what drives these changes visuomotor feedback gains in learning, by separating the effects of internal model uncertainty from visual error signal through removal of visual error information. Removing visual error information suppresses the visuomotor feedback gains in all conditions, but the pattern of modulation throughout adaptation is unaffected. Moreover, we find increased muscle co-contraction in both abrupt and gradual adaptation protocols, demonstrating that visuomotor feedback responses are independent from the level of co-contraction. Our result suggests that visual feedback benefits motor adaptation tasks through higher visuomotor feedback gains, but when it is not available participants adapt at a similar rate through increased co-contraction. We have demonstrated a direct connection between learning and predictive visuomotor feedback gains, independent from visual error signals. This further supports our hypothesis that internal model uncertainty drives initial increases in feedback gains. 
% \href{http://journals.plos.org/ploscompbiol/article/metrics?id=10.1371/journal.pcbi.1005619}{Standard Structure}

% Please include a maximum of seven keywords
\end{abstract}

\section{Introduction}

Humans constantly interact with their external world, picking up objects and manipulating them; requiring fine control of our sensorimotor system. Even simple tasks that are a part of our daily activities -- drinking a beer or peeling an apple -- involve predictive compensation for changing dynamics, such as the weight of the glass or the resistance of the apple skin. When these dynamics are consistent or change predictably (such as the weight of the glass as we drink), we are able to perform the tasks skilfully, adjusting our musculoskeletal forces to perform the tasks with little error. However, when these dynamics change unpredictably, they produce errors whose size depends on the changes in the dynamics. We respond to these errors with rapid changes in both feedback gains \cite{cluff_rapid_2013, franklin_visuomotor_2012, coltman_time_2020} and co-contraction \cite{thoroughman_electromyographic_1999, osu_short-_2002, milner_impedance_2005, franklin_adaptation_2003} in order to reduce the errors both within the current \cite{braun_learning_2009, crevecoeur_feedback_2020} and subsequent movements \cite{franklin_visuomotor_2012, franklin_feedback_2021, albert_neural_2016} through increased stiffness (intrinsic and reflexive). Simultaneously, the sensory feedback is used to update our internal model to update and gradually refine our control, which can reduce errors in subsequent movements as the predictive control is learned. That is, humans adapt to changes in dynamics by either modifying an existing motor memory (or internal model) or selecting a new motor memory \cite{oh_minimizing_2019}, which can then be used to compensate for the dynamics in a predictive manner\cite{lackner_rapid_1994, shadmehr_adaptive_1994, conditt_motor_1997}. This motor memory is not only comprised of changes in predictive force through feedforward changes in reciprocal muscle activation but also includes modified feedback gains \cite{franklin_visuomotor_2012, cluff_rapid_2013, franklin_feedback_2021} and co-contraction \cite{franklin_adaptation_2003, franklin_endpoint_2007}. Once the motor memory is learned, the co-contraction and feedback gains are gradually decreased, but often remain higher than in the null environment suggested tuning to these novel dynamics \cite{darainy_muscle_2008, franklin_visuomotor_2012, franklin_feedback_2021, franklin_rapid_2017}. 

Adaptation to novel dynamics has been shown to produce rapid increases in visuomotor \cite{franklin_feedback_2021, franklin_visuomotor_2012} and long latency stretch \cite{coltman_time_2020} feedback responses that gradually reduce with learning. We previously suggested that there are two major components to these changes in feedback responses: reactive and predictive feedback gains \cite{franklin_rapid_2017, franklin_feedback_2021}. Reactive feedback responses are up-regulated with increased internal model uncertainty \cite{franklin_feedback_2021, franklin_visuomotor_2012}. That is, when large errors signal that the internal model does not accurately predict the internal or external environmental dynamics, feedback gains are immediately up-regulated, and parallel the increases in muscle co-contraction \cite{franklin_feedback_2021, maurus2023nervous}. Predictive feedback gains on the other hand are slowly tuned according to the current dynamics and task \cite{franklin_rapid_2017, franklin_feedback_2021}. When a novel force field is applied gradually over many trials, there are no large kinematic errors, and the visuomotor feedback gains only slowly increase as adaptation occurs \cite{franklin_feedback_2021}. This slow increase in feedback gains is similar to that seen by the long latency of the stretch reflex for movement directions that were not perturbed by a carefully designed force field \cite{cluff_rapid_2013}. Overall, the pattern of feedback gains during adaptation appears to be explained by these two major components: reactive and predictive.

If we are disturbed by a sudden change in dynamics (such as when a force field is applied or changed), we see rapid increases in co-contraction, stretch reflexes, and visuomotor feedback gains, that all act to limit the disturbance. Previously we examined whether the initial reactive feedback gain is driven by internal model uncertainty \cite{franklin_feedback_2021} by contrasting the changes in visuomotor feedback gains \cite{brenner_fast_2003, sarlegna_target_2003, saunders_humans_2003, franklin_specificity_2008, knill_flexible_2011} during adaptation to abrupt and gradual changes in dynamics \cite{malfait_is_2004, klassen_learning_2005, kluzik_reach_2008, huang_persistence_2009, orban_de_xivry_contributions_2011, pekny_protection_2011, milner_different_2018}. Whereas large errors during abrupt adaptation should increase internal model uncertainty, small errors during gradual adaptation should have little effect on this measure. As predicted, we found almost no initial increase in visuomotor feedback gains during gradual adaptation to force fields, and only a slow build-up paralleling the predictive force field adaptation. However, despite this support for our theory, an alternative explanation is that simply the presence or absence of large visual errors drove these changes. Therefore, we test this possibility by having participants adapt to both abrupt and gradual curl force fields using an identical experimental design to our previous work, but where all lateral visual errors of the force field were removed using a visual clamp \cite{scheidt_interaction_2005}.

Despite the fact that visual feedback alone can drive adaptation to novel dynamics \cite{sarlegna_force-field_2010}, adaptation to novel dynamics occurs naturally in both congenitally blind individuals \cite{dizio_congenitally_2000} and in experimental conditions where the lateral visual error is removed \cite{scheidt_interaction_2005, tong_kinematics_2002, franklin_visual_2007, arce_differences_2009}. In fact, adaptation occurs at a similar rate regardless of the presence or absence of this online visual feedback \cite{tong_kinematics_2002, franklin_visual_2007}. Although some studies have suggested a slightly slower rate of adaptation without vision \cite{batcho_impact_2016}, the main effect appears to be the change in the path straightness after adaptation and increased path variability \cite{arce_differences_2009, batcho_impact_2016}. However, it is not clear what effect the removal of visual error information would have on the regulation of visuomotor feedback responses during this adaptation process. Visuomotor feedback responses are motor responses specifically elicited by visual error signals. Therefore, the removal of visual error signals throughout the experiment (and specifically the force field exposure phase) might limit any modulation of these feedback gains, since these gains would have no effect on the learning process. To examine this, we measure visuomotor feedback responses, muscle activation, and learning metrics throughout abrupt and gradual adaptation to force fields.
As we have proposed that it is internal model uncertainty that drives the changes in reactive feedback gains, we predict modulation of the visuomotor feedback gains throughout adaptation, even in the absence of visual error. However, as visual feedback can no longer be used to correct errors that build up over the movement \cite{franklin_visual_2007}, we expect that participants will rely on other control processes to reduce error and limit the effects of noise. Specifically we predict increases in muscle co-contraction which can limit the effect of motor noise \cite{osu_short-_2002, gribble_role_2003, selen_can_2005, selen_impedance_2009}. On the other hand, if it is only visual error information that drives feedback modulation during learning, we will find consistent visuomotor feedback gains throughout the adaptation and no difference between abrupt and gradual adaptation in the initial stages of learning. Therefore by contrasting the modulation of visuomotor feedback gains with and without visual error information we can distinguish between these hypotheses.

\section{Material and Methods}

Our previous experiment \cite{franklin_feedback_2021} examined the changes in the feedback gains to abrupt and gradual changes in dynamics while the reaching errors during movements were presented both physically and visually to the participants. The results showed strong modulation of the visuomotor feedback gains throughout the adaption process, which we interpreted as responses to increased or decreased uncertainty in the internal model (reactive feedback gains) and as learned feedback responses appropriate for the environment (predictive feedback gains). One major question was to what degree the presence of online visual feedback is necessary for either of these changes in the visuomotor feedback gains. Here we investigated this issue using the same paradigm and analysis \cite{franklin_feedback_2021}, but where participants do not receive online visual feedback of any lateral errors.   

\subsection{Experimental Setup}
\subsubsection{Participants}
Twelve participants participated in the experiment (5 male and 7 female: aged 25.2$\pm$ 4.6, mean $\pm$ SD)). All participants were right-handed according to the Edinburgh handedness inventory \cite{oldfield_assessment_1971} with no reported neurological disorders. Participants provided written informed consent, and the institutional ethics committee approved the experiments. 
\subsubsection{Apparatus}
Participants grasped the handle of the vBOT robotic manipulandum \cite{howard_modular_2009} with their forearm supported against gravity with an air sled (Fig. 1A). The robotic manipulandum both generated the environmental dynamics (null field, force field or mechanical channel), and measured the participants’ behavior. Position and force data were sampled at 1KHz. Endpoint forces at the handle were measured using an ATI Nano 25 6-axis force-torque transducer (ATI Industrial Automation, NC, USA). The position of the vBOT handle was calculated from joint-position sensors (58SA; IED) on the motor axes. Visual feedback was provided using a computer monitor mounted above the vBOT and projected veridically to the participant via a mirror. This virtual reality system covers the manipulandum, arm and hand of the participant, preventing any visual information about their location. The exact time that the stimuli were presented visually to the participants was determined using the video card refresh rate and confirmed with an optical sensor to prevent a time delay. Participants performed right-handed forward reaching movements in the horizontal plane at approximately 10 cm below shoulder level.

\begin{figure}[t]
\centering
\includegraphics[width=12cm]{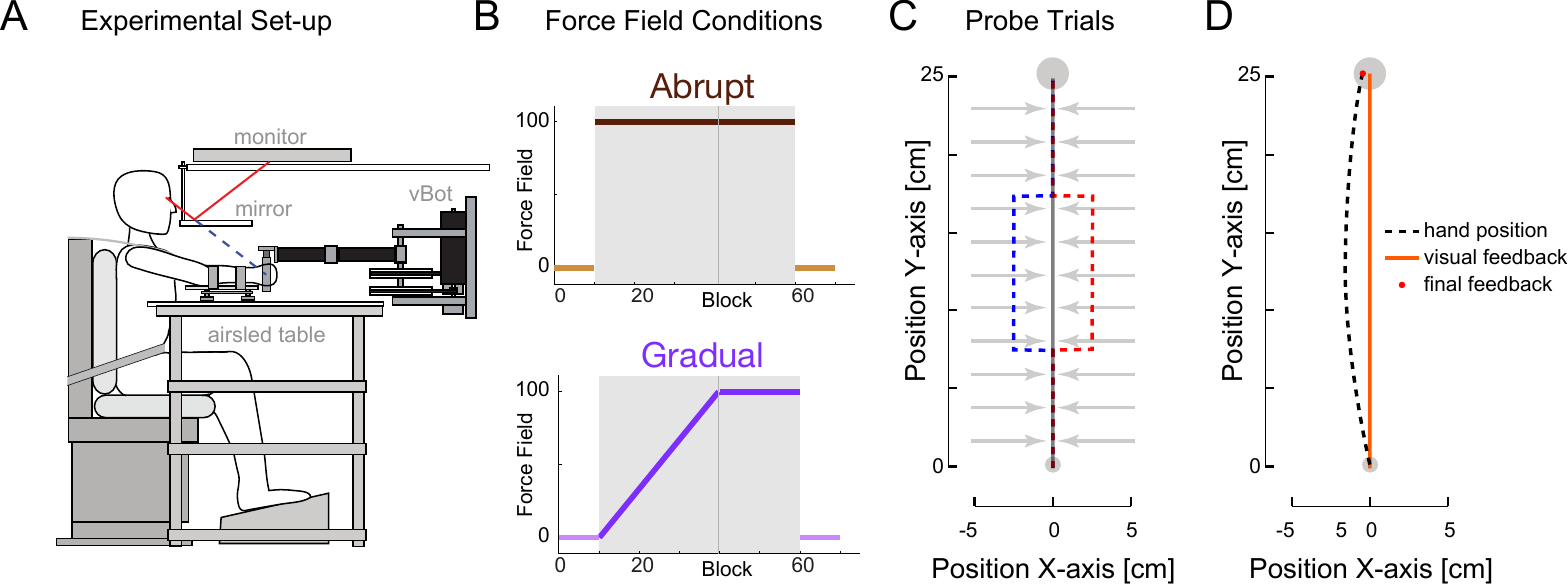}
\caption{Experimental  set-up.  A:  Participants  grasp  the  robotic  manipulandum (vBOT)  while  seated  and  visual  feedback  is  presented  veridically  using  a  top mounted  monitor  viewed  through  a  mirror  so  that  it  appears  in the  plane  of movement.  The  participant’s  forearm  is  supported  by  an  airsled.  B:  Participants experienced either an abrupt (green) or gradual (blue) onset of a velocity dependent force  field  in  different  sessions.  C:  Throughout  the  experiment,  on  random  trials visual perturbations (probe trials) were used to examine the magnitude of the visually induced motor response. During these trials, the hand (dark grey line) was physically constrained  to  the  straight  path  between  the  initial  starting position and  the final target using a mechanical channel, such that any force produced in response to the visual perturbation (blue or red dotted lines) can be measured against the virtual channel wall (grey arrows) using the force sensor. D: In all trials except the probe trials, the hand cursor (orange line) is shown to move perfectly straight to the target (visual clamp), where only the forward motion (y-axis) matches the participant's movement (black dotted line). After the trial is completed the final static location of the hand (x- and y-axis location) is presented to the participant (red circle).}
\end{figure}

\subsubsection{Setup}
Participants were seated with their shoulders restrained against the back of a chair by a shoulder harness. Movements were made from a 1.0 cm diameter start circle centered approximately 28.0 cm in front of the participant to a 2.0 cm diameter target circle centered 25 cm in front of the start circle. The participant’s arm was hidden from view by the virtual reality visual system, on which the start and target circles as well as a 0.6 cm diameter cursor used to track instantaneous hand position were projected. Participants were instructed to perform successful movements to complete the experiment. A successful movement required the hand cursor to enter the target (without overshooting) within 700 $\pm$ 75 ms of movement initiation. Overshoot was defined as movements that exceeded the target in the direction of movement. When participants performed a successful movement they were provided with feedback as how close they were to the ideal movement time of 700ms (‘great’ or ‘good’) and a counter increased by one. Similarly, when they performed unsuccessful movements they were provided with feedback as to why the movement was not considered successful (“too fast”, “too slow” or “overshot target”). Trials were self-paced; participants initiated a trial by moving the hand cursor into the start circle and holding it within the target for 1000 ms. A beep then indicated that the participants could begin the movement to the target. The duration of the movement was determined from the time that the participants exited the starting position until the time that participants entered the target.
\subsubsection{Electromyography}
Surface electromyography (EMG) was recorded from two mono-articular shoulder muscles (pectoralis major and posterior deltoid), two bi-articular muscles (biceps brachii and long head of the triceps) and two mono-articular elbow muscles (brachioradialis, and lateral head of the triceps). The EMG was recorded using the Delsys Bagnoli (DE-2.1 Single Differential Electrodes) electromyography system (Boston, MA). The electrode locations were chosen to maximize the signal from a particular muscle while avoiding cross-talk from other muscles. The skin was cleaned with alcohol and prepared by rubbing an abrasive gel into the skin. This was removed with a dry cotton pad and the gelled electrodes were secured to the skin using double-sided tape. The EMG signals were analog band-pass filtered between 20 and 450 Hz (in the Delsys Bagnoli EMG system) and then sampled at 2.0 kHz. 

\subsubsection{Probe trials to measure reflex gain}
In order to assess reflex magnitude, visually induced motor responses were examined using perturbations of the visual system similar to those previously described \cite{franklin_specificity_2008, franklin_visuomotor_2012, franklin_fractionation_2014, franklin_temporal_2016} throughout the experiments. On random trials, in the middle of a movement to the target, the cursor representing the hand position was jumped perpendicular to the direction of the movement (either to the left or to the right) by 2 cm for 250 ms and then returned to the true hand position for the rest of the movement (Fig. 1C). During these trials, the hand was physically constrained to the straight path between the starting position and the target using a mechanical channel, such that any force produced in response to the visual perturbation could be measured against the channel wall using the force sensor. The mechanical wall of the channel was implemented as a stiffness of 5,000 N/m and damping of 2Nm\begin{math}^{-1} \end{math}s for any movement lateral to the straight line joining the starting location and the middle of the target \cite{scheidt_persistence_2000, milner_impedance_2005}. As this visual perturbation was transitory, returning to the actual hand trajectory, participants were not required to respond to this visual perturbation to produce a successful trial. These visual perturbations were applied perpendicular to the direction of the movement (either to the left or the right). For comparison, a zero-perturbation trial was also included in which the hand was held to a straight-line trajectory to the target, but the visual cursor remained at the hand position throughout the trial. The onset of the displacements occurred at 7.5 cm (30\% of the length of the movement). The perturbation trials were randomly applied during movements in a blocked fashion such that one of each of the three perturbations were applied within a block of twelve trials.

\subsection{Experimental Paradigm}
This experiment examined the role that removing online visual feedback, in particular the lateral error information, plays in the adaptation process, and specifically in the control and regulation of the visuomotor feedback gains. In order to observe the effect of this sensory information, we have removed the visual feedback of the lateral error to participants throughout the experiment. Throughout this experiment, while participants could not see their hand itself, a visual cursor (yellow circle) was used to display the participant’s hand location. However all non-probe trials were visual error clamp trials \cite{scheidt_interaction_2005} in which the cursor only accurately displayed the participant’s y-position (the extent of the movement), whereas the x-position was always displayed along the straight movement to the target (x=0) (Fig 1D, right). In order to signal to the participant that their hand reached the target successfully, the target color changed from gray to white once the subject’s hand was located within 0.5 cm of the target. Participants are instructed to move the cursor into the target and stop within the target without overshooting, which requires them to use the visual information of the cursor to perform successful movements. Finally, at the end of each movement (400 ms after entering target) final endpoint feedback was provided as a stationary red dot, which subjects could use on future trials.

\subsubsection{Experimental Protocol}
Each participant performed two sessions, which were separated by a short break, in a single day. On one session, a dynamical force field was abruptly introduced (abrupt), while on the other session a directionally opposite force field was gradually applied (gradual). The order of both sessions and the force field directions were counterbalanced across participants. EMG electrodes remained in place throughout the entire experiment. Before each session, participants performed a practice of 61 null field trials in order to familiarize themselves with the movement criteria. Throughout the experiment, trials were arranged in blocks consisting of twelve trials; of which three were probe trials (visual perturbation and mechanical channel) and nine were normal reaching trials (null field or force field depending on the phase of the experiment). These probe trials were used to assess both the visuomotor feedback gain and the degree of learned force compensation. While lateral movement in the random probe trials was constrained by the mechanical channel, participants were free to move in any direction during all other trials.
Each session consisted of 4 phases. A single movement was always performed first in any new phase such that a probe trial was never the first movement. First, participants experienced the pre-exposure phase of 121 null field trials (10 blocks of twelve trials plus one initial trial). Next, the initial exposure stage consisted of 361 force field trials (30 blocks plus one initial trial). In the abrupt condition, the full force field was applied from the first trial, whereas in the gradual condition the force field was scaled up from one trial to the next over the 361 trials (Fig. 1B). The force field was a velocity-dependent curl force field where the force in N (Fx, Fy) on the hand was computed as:
% Equations should be inserted using standard LaTeX equation and eqnarray environments, not as graphics, and should be set in the main text
% This is an equation, numbered
\begin{eqnarray}
\left[
\begin{array}{ccc} F_{x} \\ F_{y} \end{array}\right]=b\left[
\begin{array}{ccc}
0 & -1 \\
1 & 0 \\
\end{array}\
\right]\left[
\begin{array}{ccc}
\Dot{x} \\
\Dot{y}\\
\end{array}
\right]
\end{eqnarray}
% And one that is not numbered
%\begin{equation*}
%e^{i\pi}=-1
%\end{equation*}
depending on the participants' hand velocity ($\Dot{x},\Dot{y}$) [m/s] and the scaling factor $b$ which was either 0.16 N/m/s (clockwise curl force field: CW) or -0.16 N/m/s (counter-clockwise curl force field: CCW). Once the initial exposure phase was completed, both groups of participants performed the final exposure phase consisting of another 20 blocks of trials (241 trials). Finally, participants experienced the post-exposure phase in which 10 blocks (121 trials) of null field trials were performed. Participants were required to take short breaks every 200 movements throughout the experiment. They were also allowed to rest at any point they wished by releasing a safety switch on the handle. 
\subsection{Analysis}
Analysis of the experimental data was performed using Matlab R2019a. EMG data were band-pass filtered (30 – 500Hz) with a tenth-order, zero phase-lag Butterworth filter and then rectified. Position, velocity and endpoint force were low-pass filtered at 40 Hz with a fifth-order, zero phase-lag Butterworth filter. Statistics were performed in Matlab and JASP 0.14.1 \cite{JASP2020}. Statistical significance was considered at the p<0.05 level for all statistical tests.
\subsubsection{Hand Path Error}
The maximum perpendicular error (MPE) was used as a measure of the straightness of the hand trajectory. On each trial, the MPE is the maximum distance on the actual trajectory that the hand reaches perpendicular to the straight-line path joining the start and end targets (errors to the left are defined as negative and errors to the right are defined as positive). The MPE was calculated for each non-probe trial throughout the learning experiment.

\subsubsection{Force Compensation}
In order to examine the predictive forces exerted by the participants throughout the experiment, the forces against the mechanical channel walls on the probe trials were used. On each trial, the amount of force field compensation was calculated by linear regression of the measured lateral force against the channel wall onto the ideal force profile required for full force field compensation \cite{smith_interacting_2006}. Specifically, the slope of the linear regression through zero is used as a measure of force compensation. The ideal force field compensation was estimated as the product of the y-velocity and the force field scaling factor. In the null field, the ideal force field compensation is based upon the compensation required in the curl force field \cite{howard_gone_2012}. Therefore, values in the null force field before learning (pre-exposure phase) should be close to zero. In the gradual curl force field, the force compensation on a given trial was calculated based on the current strength of the force field at this specific trial.

\subsubsection{Electromyographic Activity}
For plotting purposes, the EMG was adjusted to the mean value of EMG in the null field trials prior to the force field exposure for each muscle of each participant prior to averaging. To examine differences in the overall muscle activity across the experiment, the integral of the rectified EMG data was taken over 700 ms from 100 ms before movement start until 600 ms after movement start. The EMG data was averaged across all trials in a block. The EMG was normalized and then averaged across participants (but not across muscles). To normalize for a particular muscle, a single scalar was calculated for each participant and used to scale the muscle’s EMG traces for all trials for that participant. The scalar was chosen so that the mean (across trials) of the EMG data averaged over the whole experiment was equal across participants (and set to the mean over all the participants). This puts each participant on an equal scale to influence any response seen in the data. For comparison across muscles, the EMG values were further scaled for each muscle relative to the mean value in the pre-exposure phase (expressed as a percentage value relative to the pre-exposure activity). The muscle activity was further separated into the amount related to co-contraction and the amount related to force production for each of the three muscle pairs. In order to examine the amount of co-contraction, the minimum value of EMG between the two muscles making up a muscle pair was determined and multiplied by 2 as both muscles would contribute to the increased stiffness. The activation that would correspond to the change in force was determined as the maximum muscle activity of the two muscles of the muscle pair subtracted by the minimum of the two muscle activities. Differences in these measures across the gradual and abrupt conditions were examined using a t-test in Matlab.

\subsubsection{Rapid Visuomotor Responses}
Individual probe trials were aligned on visual perturbation onset. The response to the right visual perturbation on probe trials was subtracted from the response to the left perturbation on probe trials in order to provide a single estimate of the motor response to the visual perturbation for each block. To examine the feedback gain, we calculated the average post-perturbation force over two intervals: the first corresponding to an rapid involuntary response (180–230 ms) \cite{franklin_specificity_2008}, and the second to a slower response (230-300 ms) \cite{franklin_fractionation_2014, franklin_rapid_2017}.
To examine the electromyographic responses to the visuomotor perturbations, the EMG traces were divided by the mean value of the muscle activity in the null force field (pre-exposure) for that muscle in that participant between -50 and 50 ms relative to the onset of the perturbation. Muscular responses were considered over two intervals: 90-120 ms and 120-180 ms as in previous studies \cite{franklin_fractionation_2014, gu_trial-by-trial_2016, cross_visual_2019, franklin_feedback_2021}.

\subsubsection{Comparisons}
In order to compare the final values after learning, the mean measures (MPE, force compensation and visuomotor feedback gain) over the last 15 blocks in the final exposure phase were contrasted between abrupt and gradual conditions using frequentist repeated measures ANOVAs with a between subjects factor of condition order using JASP 0.18.1. The initial visuomotor feedback gains in the early exposure phase were also contrasted between abrupt and gradual conditions using the first 5 blocks in the exposure phase. To complement the frequentist approach we also perform equivalent Bayesian repeated measures ANOVAs using JASP 0.18.1 \cite{JASP2020}. Here we use the Bayes Factors $(BF_{10}$) to evaluate evidence for the null hypothesis. In order to examine the after effects on the first block of trials in the post-exposure phase for both MPE and force compensation, we subtracted the mean values in the pre-exposure phase from the value of the first block in the post-exposure phase. We then performed a paired t-test (and a Bayesian paired t-test) to examine any differences between the gradual and abrupt adaptation groups. Significant differences were examined at the p<0.05 level.

\section{Results}
Participants grasped the handle of a robotic manipulandum (Fig. 1A) and performed forward reaching movements. After an initial baseline phase in a null field, participants were presented with either an abrupt or gradual introduction of a velocity-dependent force field (Fig. 1B). Visuomotor feedback gains were measured on random probe trials where a brief visual perturbation of a hand cursor (to the right or left of the movement) was applied while the physical hand was constrained to move within a mechanical channel to the target (Fig. 1C). Similar to our previous work \cite{franklin_feedback_2021}, predictive force compensation, kinematic error, and muscle activation patterns were examined over the entire experiments. However, in this study participants were not provided with online visual feedback about the lateral movement of their hand (no lateral error information was provided, Fig. 1D) at any point during the experiment.

\begin{figure}
\centering
\includegraphics[width=11cm]{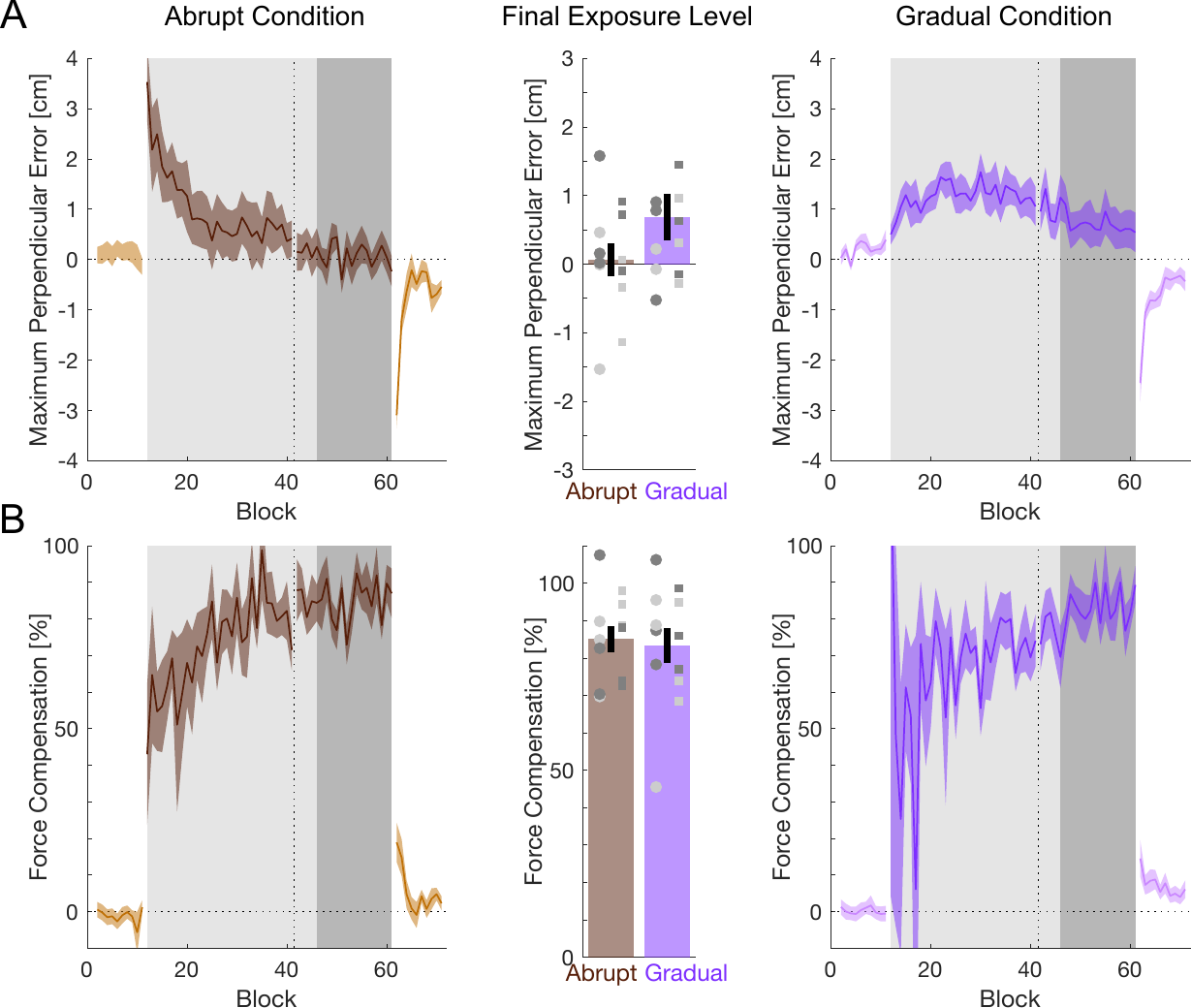}
\caption{Comparison of adaptation between abrupt (brown) and gradual (purple) exposure to the curl force fields. A: The mean (solid line) and standard error of the mean (shaded region) of the maximum perpendicular error (MPE) of the hand trajectory over the experiment. The sign of the MPE measurement from the CCW force fields was flipped so that all errors produced by the force field were shown to be positive. The gray shaded area (and darker brown and purple colors) indicate the period over which the curl force field was applied. The vertical dotted line shows the time point at which the gradual force field was the same strength as the abrupt force field. Dark gray shaded area indicates the last 15 blocks of the full force field exposure phase over which the final levels of adaptation were compared. The center bar graph compares this final level of MPE across the two force fields. The error bar represents the standard error of the mean (s.e.m.). Each participant’s final value is shown with a point. A light grey point indicates the CW force field whereas a dark grey point indicates the CCW force field. The square indicates the first force field experienced whereas the circle indicates the second time participants experienced a force field. There were no significant differences between the conditions. B: Force compensation level over the experiment as measured on the channel trials. A value of 100\% indicates perfect compensation for the force field. Force compensation in the null field was quantified with respect to the full force field value, so a value of 0 is expected. In the gradual condition, as the force field is ramped up, the force compensation is expressed as a percentage of the force field that has been applied to the participant at this point in the experiment. Values plotted as in A. Similar levels of force compensation were found for abrupt and gradual conditions with no significant difference over the last 15 blocks of exposure.}
\end{figure}

\subsection{Behavior}
All participants were exposed to curl force fields with both an abrupt and gradual introduction, but where both the order and the field direction were counterbalanced across participants. After a short pre-exposure phase (in the null field), a curl force field was applied. As expected this produced a large increase in the kinematic error in the abrupt condition (Fig. 2A left, brown) but only a small increase in the gradual condition as the strength of the force field got close to the final level (Fig. 2A right, purple). While the MPE reduced over the final blocks in both abrupt and gradual conditions, it appeared that there was slightly larger MPE for gradual compared to abrupt condition in the final level of kinematic error. However, a comparison using a repeated measures ANOVA of the last 15 blocks in the final exposure phase (dark gray shaded area) failed to find a significant difference in the MPE between the abrupt and gradual conditions ($F_{1, 10}=2.779, p=0.126; BF_{10}=1.165$) (Fig. 2A center bar graph). There was also no between subjects effect of condition order ($F_{1,10}=0.091, p=0.769; BF_{10}=0.483$). When the force field was removed (post-exposure phase) there was a large initial increase in MPE in the opposite direction to that of the force field for both groups (Fig. 2A), but this reduced quickly. There was no significant difference between the kinematic after effects on the first block of trials between the abrupt and gradual adaptation groups ($t_{11}=1.21, p=0.251$; $BF_{10}=0.524$).

On random trials, a mechanical channel constrained participants’ hand to move in a straight line to the target. On these trials, the predictive force compensation, indicating the percent of perfect adaptation to the force field, can be estimated (Fig. 2B). Importantly in the gradual condition, the percentage adaptation is expressed as a function of the current level of force field during the ramp phase. In both conditions, participants demonstrated over 80\% force compensation after reaching 100\% strength of the force field (blocks 40-60). When comparing the final levels of adaptation in the abrupt and gradual conditions (final 15 blocks of exposure phase), we again found no significant difference (dark gray shaded area in the figures) in the force compensation (Fig. 2B center) between the abrupt and the gradual conditions ($F_{1,10}=0.094, p=0.765; BF_{10}=0.385$) and no between subjects effect of condition order ($F_{1,10}=0.014, p=0.910; BF_{10}=0.488$). Therefore, participants in the abrupt condition adapted similarly to the force field, although they had a much greater number of trials in which they were presented with the full magnitude of the force field. In the post- exposure phase, there was also no significant difference between the force compensation after effects on the first block of trials between the abrupt and gradual adaptation groups ($t_{11}=0.729, p=0.481$; $BF_{10}=0.361$).

\subsection{Muscle Activity}
Muscle activity (sEMG) recorded from six muscles during the abrupt (brown) and gradual (purple) conditions are shown for both the counter-clockwise (Fig. 3A) and clockwise (Fig. 3B) force fields. EMG values have been normalized to the null field level for all muscles, and are shown here only on the trials in which a mechanical channel is applied such that there is no force field and no trajectory errors. In the abrupt condition, large increases in muscle activation initially occurred in all six muscles, which slowly reduced as the learning occurred (brown), while only slow increases in muscle activity occurred in the gradual condition (purple). In both cases, the initial increases in co-contraction reduced throughout the exposure period, and increased final levels of muscle activation primarily occurred in the muscles compensating for the force fields. That is, posterior deltoid and triceps lateralis in the counter-clockwise curl force field, and pectoralis major and biceps brachii in the clockwise curl force field. These changes in muscle activation changes are as expected for adaptation to the CW and CCW curl force fields where increased flexor and extensor torques are required respectively. When the force field was suddenly removed (after block 60), there were further increases of muscle activity, especially in the antagonist muscles – the muscles that were not acting to compensate for the force fields (e.g. posterior deltoid and triceps longus in the CW force field for a forward movement). This suggests a brief increase in co-contraction when the force field is removed and increased kinematic errors occur (initial after effects). As the muscle activity on this figure only includes movements with a mechanical channel, the expectation is that most activity is pre-planned and not a reaction to kinematic errors experienced on the specific trial.

\begin{figure}[t]
\centering
\includegraphics[width=12cm]{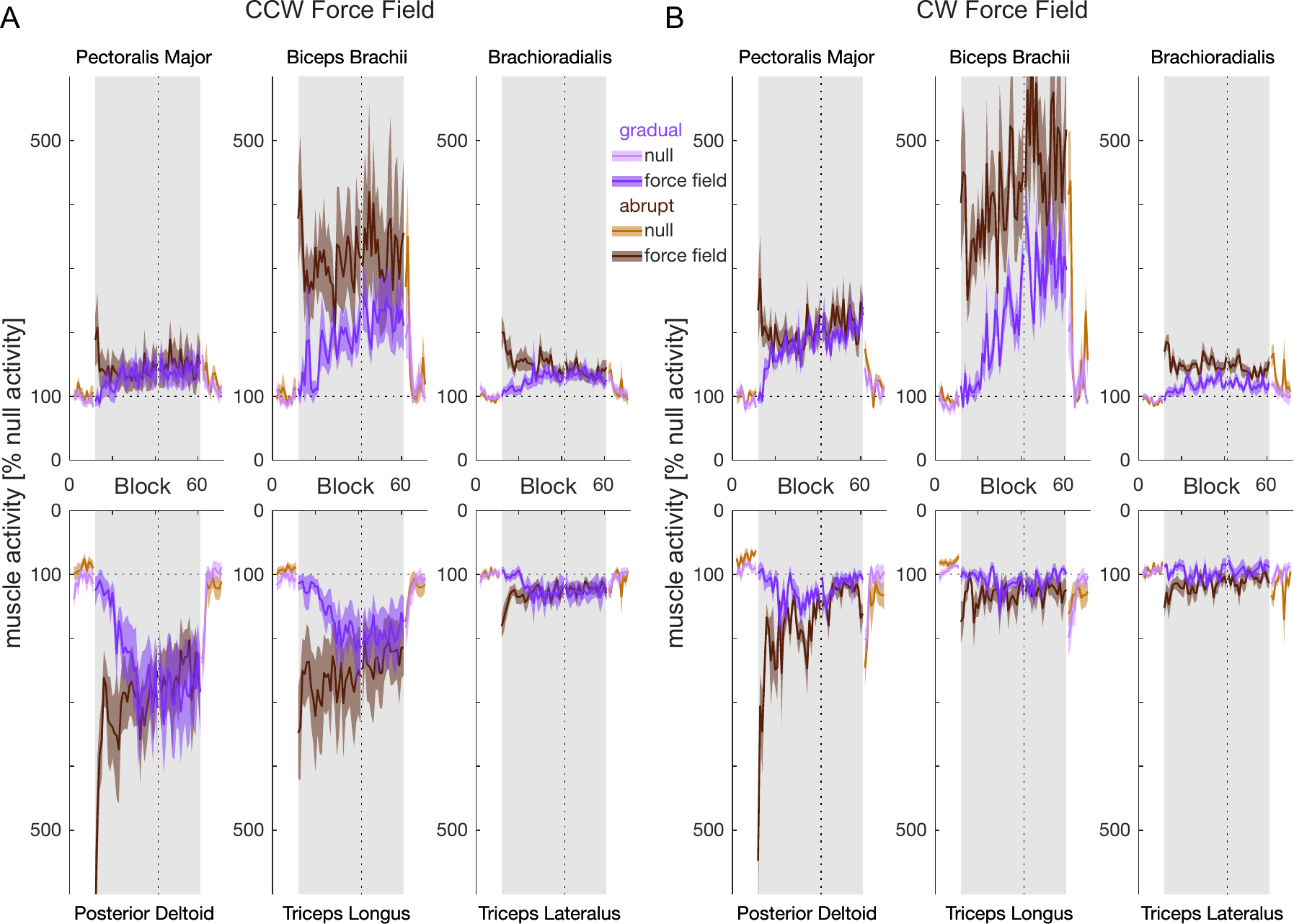}
\caption{Comparison of muscle activity (EMG) across the experiment for abrupt (brown) and gradual (purple) exposure to curl force fields. Light brown and purple indicate values in the null field whereas dark brown and purple indicate values in the curl force field. Data is only from the randomly interspersed mechanical channel trials where the curl force field was not applied. Increasing extensor muscle activity is plotted in the downward direction such that increasing co-contraction results in larger distances between muscle pairs (shoulder, bi-articular, elbow). A: Muscle activity (mean and s.e.m) during adaptation to the CCW curl force field. Muscle activity was normalized for each of the six muscles to the mean value in the pre-exposure phase before averaging across participants. EMG values were calculated as the integrated muscle activity from -100 to 600 ms relative to the start of the movement. Grey shaded region indicates the exposure phase and the vertical dotted line indicates the point at which the gradual force field is equal to the full force field value. B: EMG in the CW curl force field. Across both force fields, high levels of muscle activation were initially observed when the force field was introduced abruptly (brown) but not when introduced gradually (purple).}
\end{figure}

The temporal profile of muscle activity after adaptation to abrupt or gradual dynamics was examined for both the CCW and CW force fields (Fig. 4) with null field trials shown for comparison. As expected, the muscle activity profile in the null field prior to either abrupt or gradual force field exposure is similar. After adaptation, we also find similar temporal profiles of muscle activation for most muscles, but some differences between the conditions. Our previous research \cite{franklin_feedback_2021} suggested slightly larger increases in biarticular muscles under gradual exposure to force fields and larger should joint activation for abrupt force fields. However, there was no evidence supporting this finding in the current results. Here, if anything, the final level of adaptation in the abrupt condition was slightly larger in the biarticular muscles, although the error bars clearly overlap throughout. It is unclear whether this difference with previous results is due to the absence of visual error information in the current experiment or whether there is no such effect on final adaptation by differences in the exposure rate.

\begin{figure}[t]
\centering
\includegraphics[width=12cm]{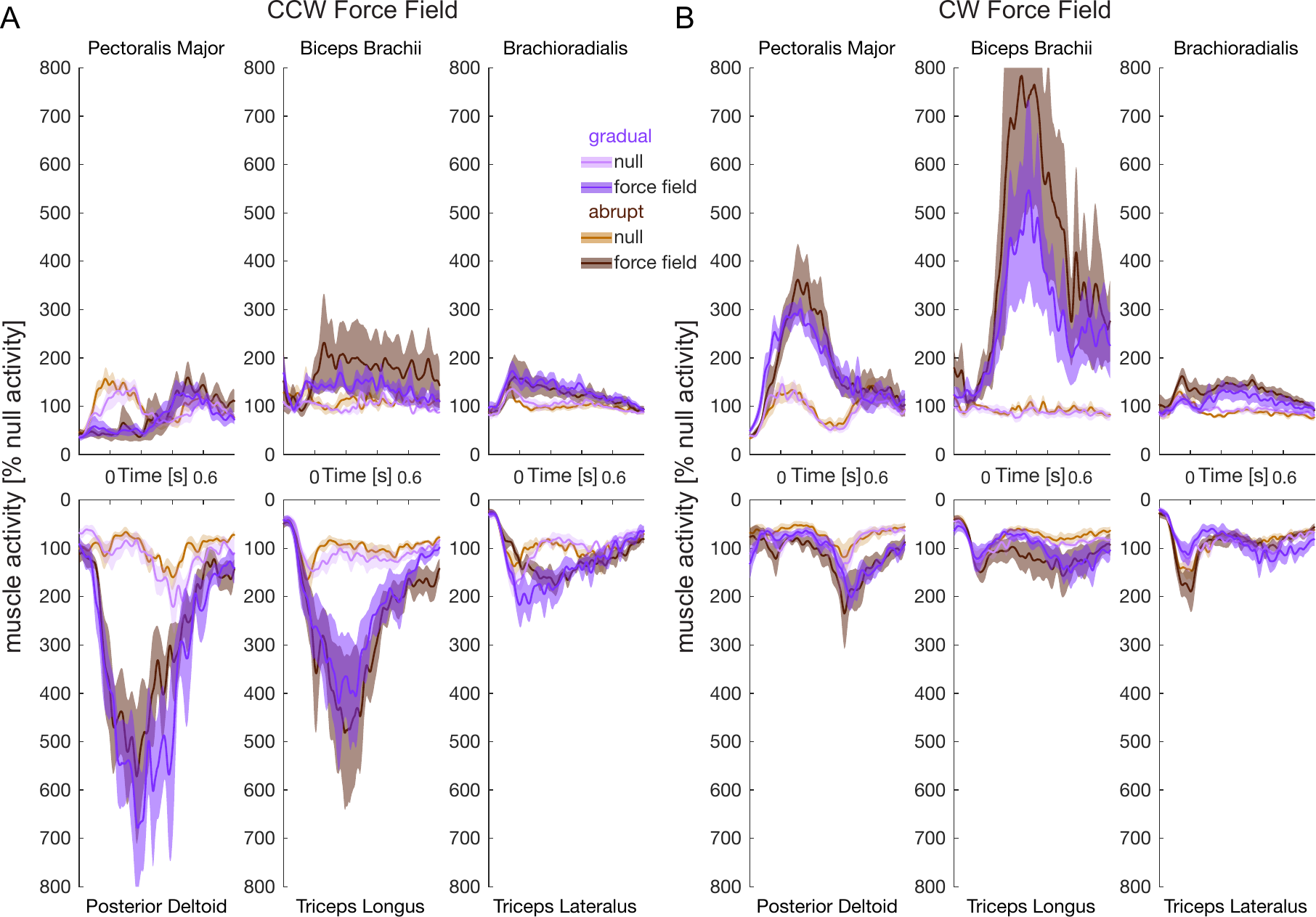}
\caption{Temporal profiles of muscle activity in the null force field and after adaptation for the abrupt (brown) and gradual (purple) conditions. Data is only from movements within a mechanical channel where the curl force field was not applied. Increasing extensor muscle activity is plotted in the downward direction such that increasing co-contraction results in larger distances between muscle pairs (shoulder, bi-articular, elbow). Prior to averaging across participants, the EMG has been smoothed with a 50-point (25ms) smoothing function. A: EMG profiles in the CCW curl force field. Null field activity (all 10 blocks in the pre-exposure phase) prior to adaptation is indicated by the light brown and light purple traces. Final adaptation activity (all 20 blocks in the final exposure phase) is indicated by the dark brown and dark purple traces. Muscle activity has been aligned to the start of the movement (0 s). Solid lines indicate mean across participants and shaded regions indicate s.e.m. B: EMG profiles in the CW curl force field.}
\end{figure}

When we quantify the increases in EMG for both initial and final exposure (Fig 5), we again find similar results. The abrupt condition showed large increases in muscle activation for almost all muscles (increased co-contraction) during the initial exposure (Fig. 5A), where little or no increases in muscle activity were seen in the gradual condition. However, during the final exposure phase we see a similar level of muscle activity in both abrupt and gradual conditions (Fig. 5B). To quantify the degree of co-contraction and adaptation in the experiments across the gradual and abrupt conditions, we calculated the co-contraction and adaptation indices (Fig. 5C and 5D). The co-contraction index is a simple measure to capture the relative amount of activation in antagonistic muscle pairs, whereas the adaptation index is designed to indicate the amount of muscle activity in a specific direction (reciprocal activation) that might be directed to compensate for the force field. In the initial exposure phase (Fig. 5C), both the co-contraction index ($t_{11}=3.804, p=0.003; BF_{10}=16.27$) and adaptation index ($t_{11}=4.241, p=0.001; BF_{10}=30.42$). were much higher in the abrupt than gradual conditions. However, in the final exposure phase (Fig. 5D), there were no significant differences in either the co-contraction ($t_{11}= 1.026, p = 0.327; BF_{10} = 0.446$) or adaptation ($t_{11}= 0.528, p = 0.608; BF_{10} = 0.324$) measures. Therefore, overall the final muscle activity levels were similar across both presentations of the dynamics.

\begin{figure}
\centering
\includegraphics[width=12cm]{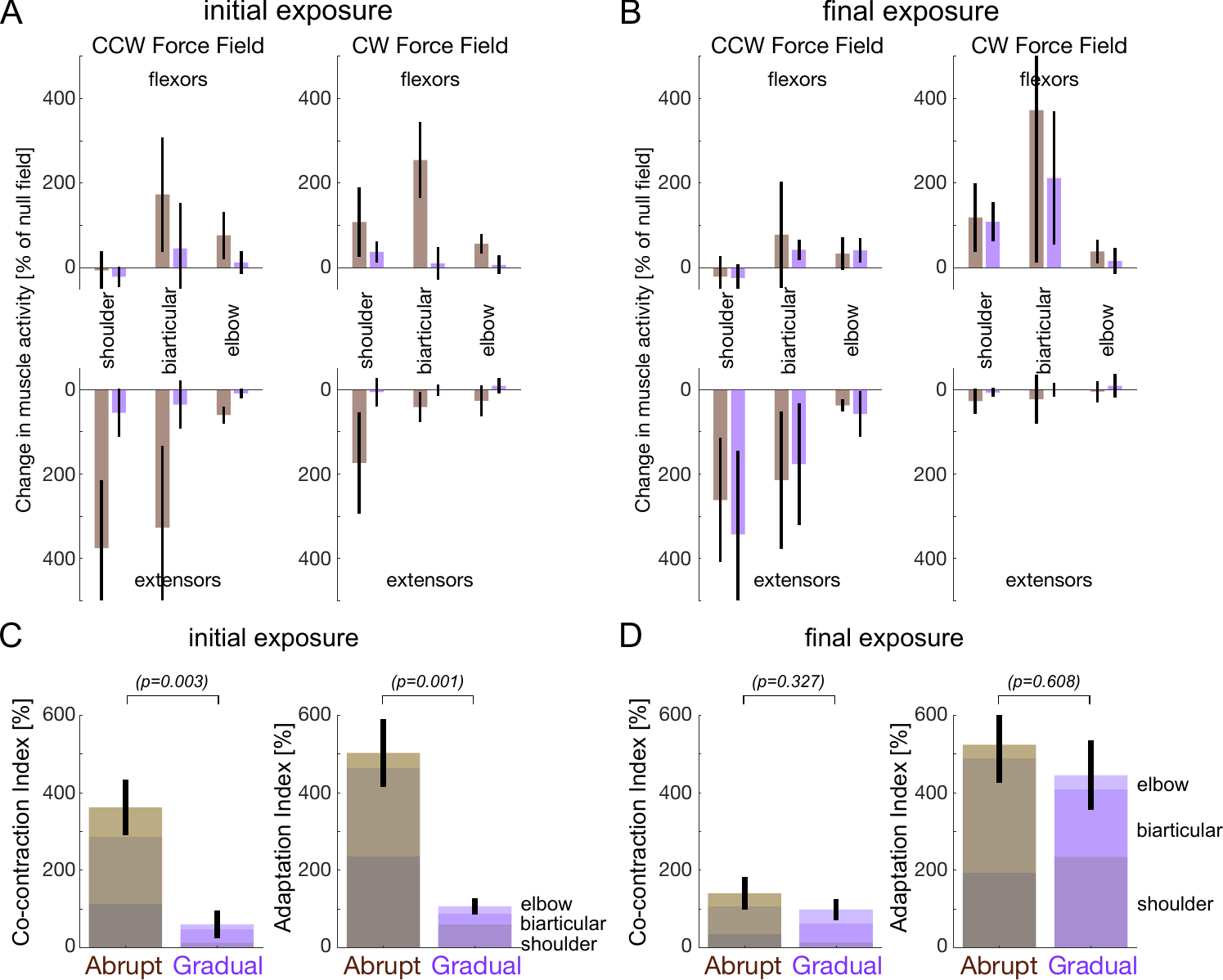}
\caption{Muscle activity in initial and final exposure to abrupt (brown) and gradual (purple) application of force fields. Data is only from movements within a mechanical channel where the curl force field was not applied. A: Initial exposure (first 10 blocks in initial exposure phase) to abrupt force field elicits large co-contraction in shoulder, biarticular and elbow muscles for both CCW and CW curl fields. Muscle activity is calculated as increase relative to the null field activity. Bars indicate mean ($\pm$ 95\% confidence intervals) integrated muscle activity from -100 to 600 ms from the start of the movement. If the error bars do not overlap then this indicates a significant difference at $p<0.05$. B: Final exposure (last 10 blocks in final exposure phase) shows similar levels of change in muscle activity for both abrupt and gradual change in dynamics. C: The co-contraction index and adaptation index of muscle activity in the initial exposure. Bar indicates total values across the muscle pairs ($\pm$ s.e.m.) and the colors indicate the relative contribution from the shoulder muscles (dark colors), biarticular muscles (medium colors) and elbow muscles (light colors). Values are across both the CCW and CW curl fields. Statistics indicate result of t-test. D: The co-contraction index and adaptation index of muscle activity in the final exposure periods.}
\end{figure}

\subsection{Visuomotor Feedback Responses}

\begin{figure}[t]
\centering
\includegraphics[width=11cm]{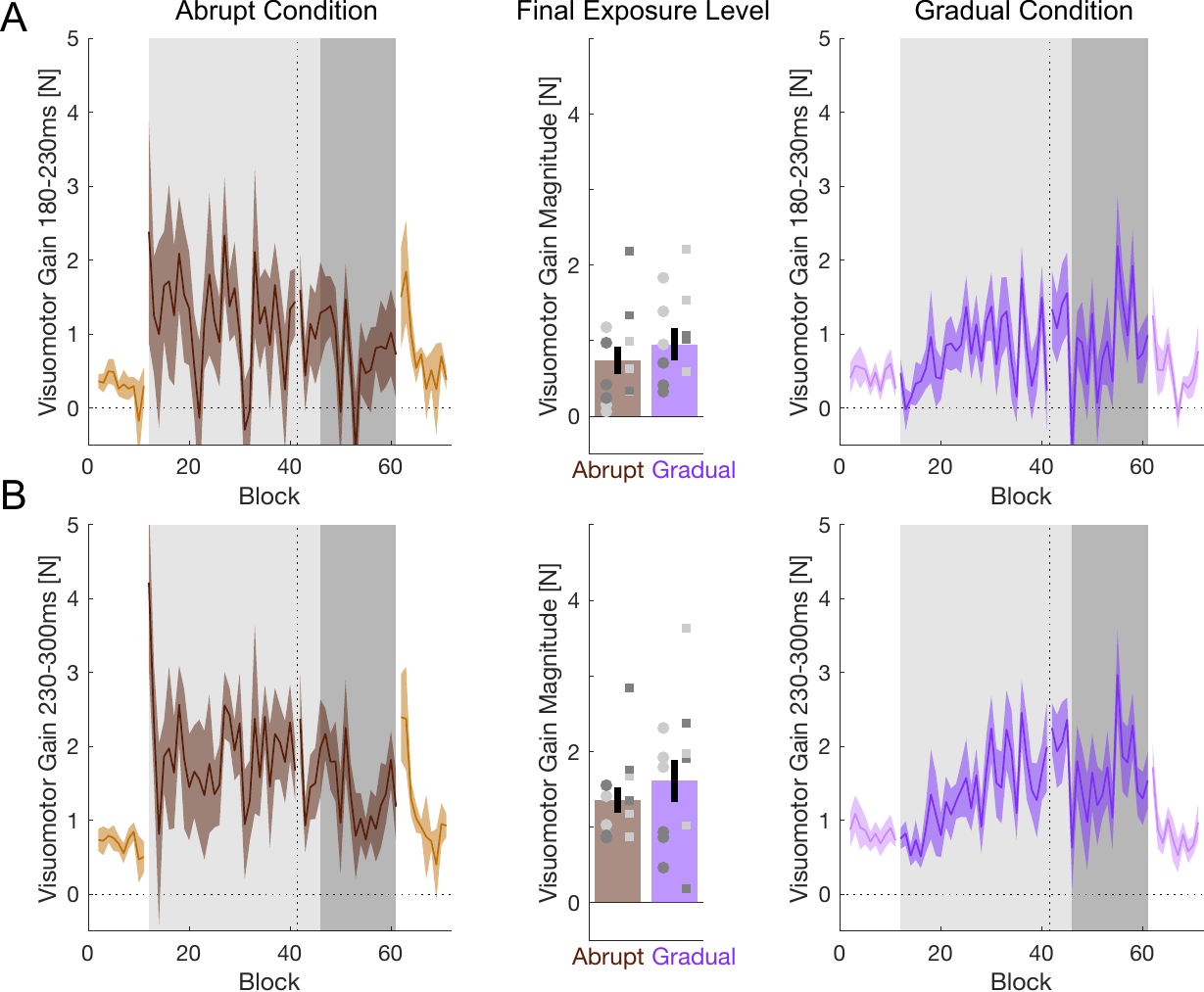}
\caption{Changes in visuomotor feedback gain during adaptation to abrupt (brown) and gradual (purple) curl force fields with no lateral visual error presented to participants. A: Visuomotor feedback gains during 180-230ms after the visual perturbation onset. Figure plotted as in Fig. 2. B: Visuomotor feedback gains during 230-300ms after the visual perturbation onset. Visuomotor feedback gains are measured on probe trials within a mechanical channel where curl force fields were not applied.}
\end{figure}

Visuomotor feedback gains were measured in both the abrupt and gradual conditions using probe trials throughout exposure to the curl force fields. The visuomotor feedback response was quantified over both an early interval (180-230 ms) and late interval (230-300 ms) after perturbation onset (Fig 6B). The onset of the abrupt change in dynamics produced a rapid increase in the visuomotor gain (brown traces) which then slightly decreased over learning as seen in previous work \cite{franklin_visuomotor_2012, franklin_feedback_2021}. When the curl force field was applied gradually, the visuomotor gains were initially unchanged, but then gradually increased over the whole exposure period and finally plateaued as the full level of force field was applied in the final exposure trials as in our previous work \cite{franklin_feedback_2021}. Initial exposure to the force field produced different responses depending of the condition of the force field, with abrupt adaptation causing a much larger increase in visuomotor feedback responses in both the early ($F_{1,10}=11.115, p=0.008$) and late intervals ($F_{1,10}=10.148, p=0.010$). Bayesian statistics supported these findings, with models including the force field condition providing the highest evidence in both the early ($BF_{10}=17.885$) and late intervals ($BF_{10}=26.244$). However, the visuomotor gains in the last 15 blocks of the exposure phase were not significantly different between the abrupt and gradual conditions in either the early ($F_{1,10}=0.927, p=0.358$; Fig. 6A bar plot) or late intervals ($F_{1,10}=1.150, p=0.309$; Fig. 6B bar plot). For both cases, there was also no effect of condition order (early: $F_{1,10}=0.409, p=0.537$; late: $F_{1,10}=0.002, p=0.963$). The Bayesian statistics supported the finding of no differences across conditions, with the null model providing the highest evidence for both the early ($BF_{10}<0.632$ for all other models) and late intervals ($BF_{10}<0.548$ for all other models). That is, both abrupt and gradual adaptation resulted in the same final level of visuomotor gains despite the large difference in kinematic errors during initial exposure. When the force field was removed abruptly, we see an initially high visuomotor gain that decreased rapidly in these null field trials in both conditions. Overall these results support our theory that internal model uncertainty affects feedback gains. When internal model uncertainty is initially high (early exposure in abrupt condition) we find increased feedback gains compared to when internal model uncertainty is low (early exposure in gradual condition). However, by the end of the exposure period both conditions show similar levels of visuomotor feedback gains, suggesting similar final adaptation of the visuomotor feedback gains to the dynamics.

\begin{figure}
\centering
\includegraphics[width=9cm]{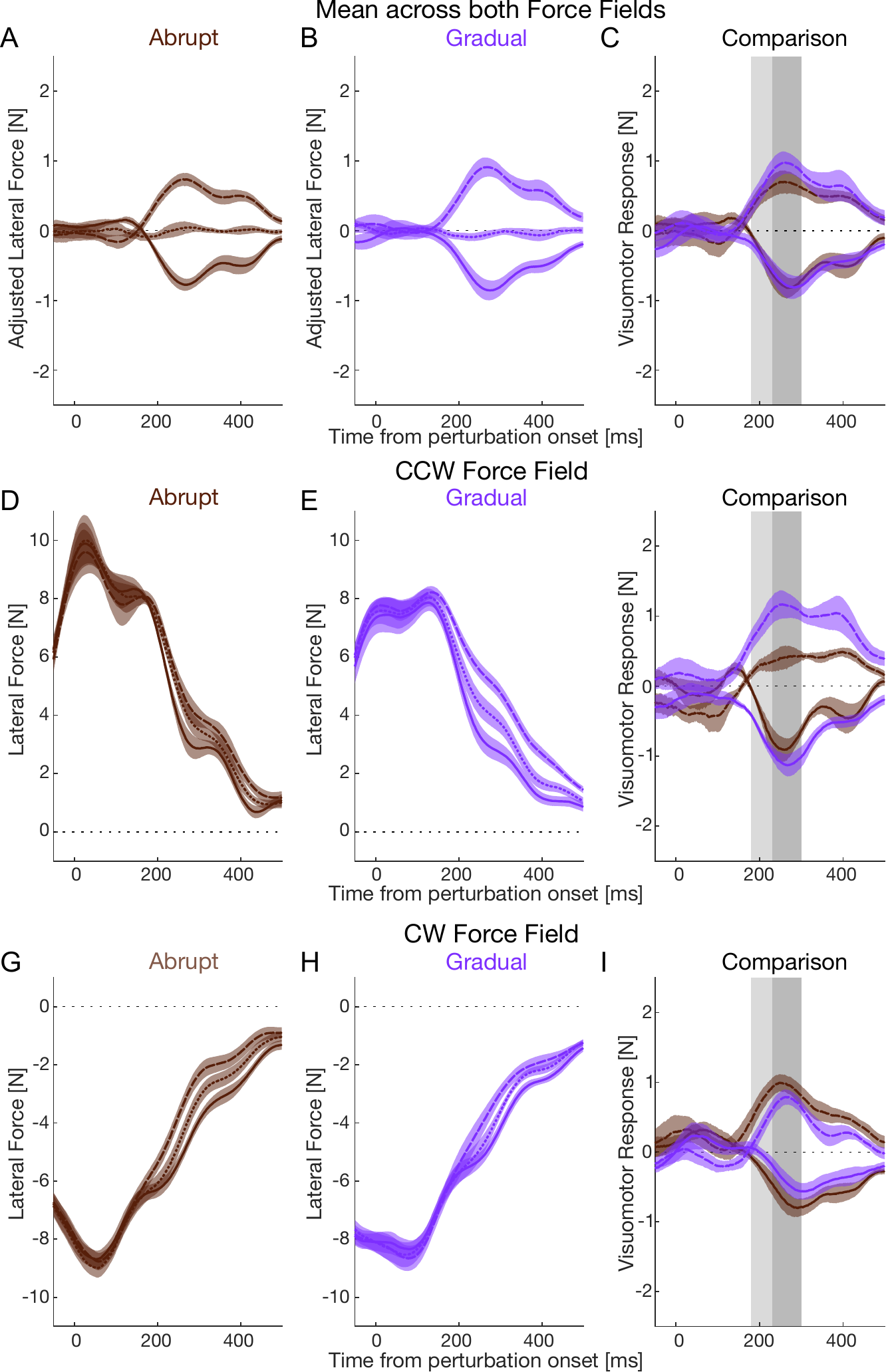}
\caption{Comparison of final visuomotor feedback responses after adaptation to the abrupt (brown) and gradual (purple) curl force fields. A: Mean ($\pm$ s.e.m.) lateral force produced after rightward (solid lines), zero (dotted lines) and leftward (dashed lines) visual perturbations across both CCW and CW fields in the abrupt condition. Lateral force was adjusted by subtracting the mean lateral force across all probe trials in each force field. B: Lateral force in response to visual perturbations across both CCW and CW fields in the gradual condition. C: Visuomotor responses (zero perturbation subtracted) across both CCW and CW fields. Light grey shaded region indicates the early visuomotor response interval (180-230 ms) while the dark grey region indicates the late visuomotor response interval (230-300 ms). D: Lateral force produced in response to visual perturbations after adaptation to the abrupt onset of the CCW force field. E: Lateral force produced in response to visual perturbations after adaptation to the gradual onset of the CCW force field. F: Visuomotor responses in the CCW force field after abrupt and gradual adaptation. G-I: Visuomotor force responses after adaptation to the CW force field.}
\end{figure}

Prior work has demonstrated that movement kinematics can influence feedback gains \cite{crevecoeur_feedback_2013, cesonis_time--target_2020, cesonis2022contextual}, so we also contrasted the peak velocities across conditions. The mean peak forward velocities ($\pm$ std) were 63.99 $\pm$ 5.05 m/s in the gradual condition and 63.90 $\pm$ 5.29 m/s in the abrupt condition. To test for differences across the adaptation period (early and late), we performed a repeated ANOVA. There was no difference between abrupt and gradual ($F_{1,11}=3.027, p=0.110$), between early and late exposure phases ($F_{1,11}=0.443, p=0.519$), or significant interaction effect ($F_{1,11}=0.475, p=0.505$). This was supported by Bayesian statistics, with the null model providing the highest evidence ($BF_{10}<0.808$ for all other models).

We further contrasted the visuomotor feedback gains at the end of the exposure period by plotting the lateral hand force of the participant against the wall of the mechanical channel in the probe trials as a function of the time from perturbation onset (Fig. 7). The visuomotor feedback force produced after the abrupt introduction of the force field (Fig. 7A) looks similar to that after the gradual introduction of the force field (Fig. 7B). After subtracting the zero-perturbation condition, the visuomotor response is similar in both conditions (Fig. 7C). As participants in these experiments adapted to both the CCW and CW force fields, we also examined the force response in each of these force fields separately (Fig. 7D-I). Although we find roughly similar lateral forces against the mechanical channel wall for both the abrupt and gradual conditions in the CCW force field, when we subtract the zero-perturbation condition it looks as though the forces are larger in the gradual condition (Fig. 7F). In the CW force field, the lateral forces are in the opposite directions, but here the comparison shows a similar response in both abrupt and gradual conditions, with the abrupt forces slightly larger (Fig. 7I). The CCW and CW force fields require opposite adaptive forces which can be clearly seen in the force traces (e.g. compare Fig. 7 D and G). In each force field a perturbation in one direction would be resisted by highly active muscles whereas in the other direction these muscles would have much lower activation (e.g. Fig. 5). This would then be reversed in the opposite force field. However, despite these differences the visuomotor force response is roughly equal in both perturbation directions. To support this claim, we compared the feedback force response to perturbations in the direction of the force field to those opposite to the direction of the force field using a t-test. We found no significant differences in either the early  ($t_{22}=0.261, p=0.796$; $BF_{10}=0.383$) or late ($t_{22}=0.567, p=0.576$; $BF_{10}=0.420$) intervals. This result supports our previous results that visuomotor feedback responses do not exhibit gain scaling \cite{franklin_visuomotor_2012, franklin_rapid_2017, franklin_feedback_2021} at the level of force responses.

\begin{figure}
\centering
\includegraphics[width=13cm]{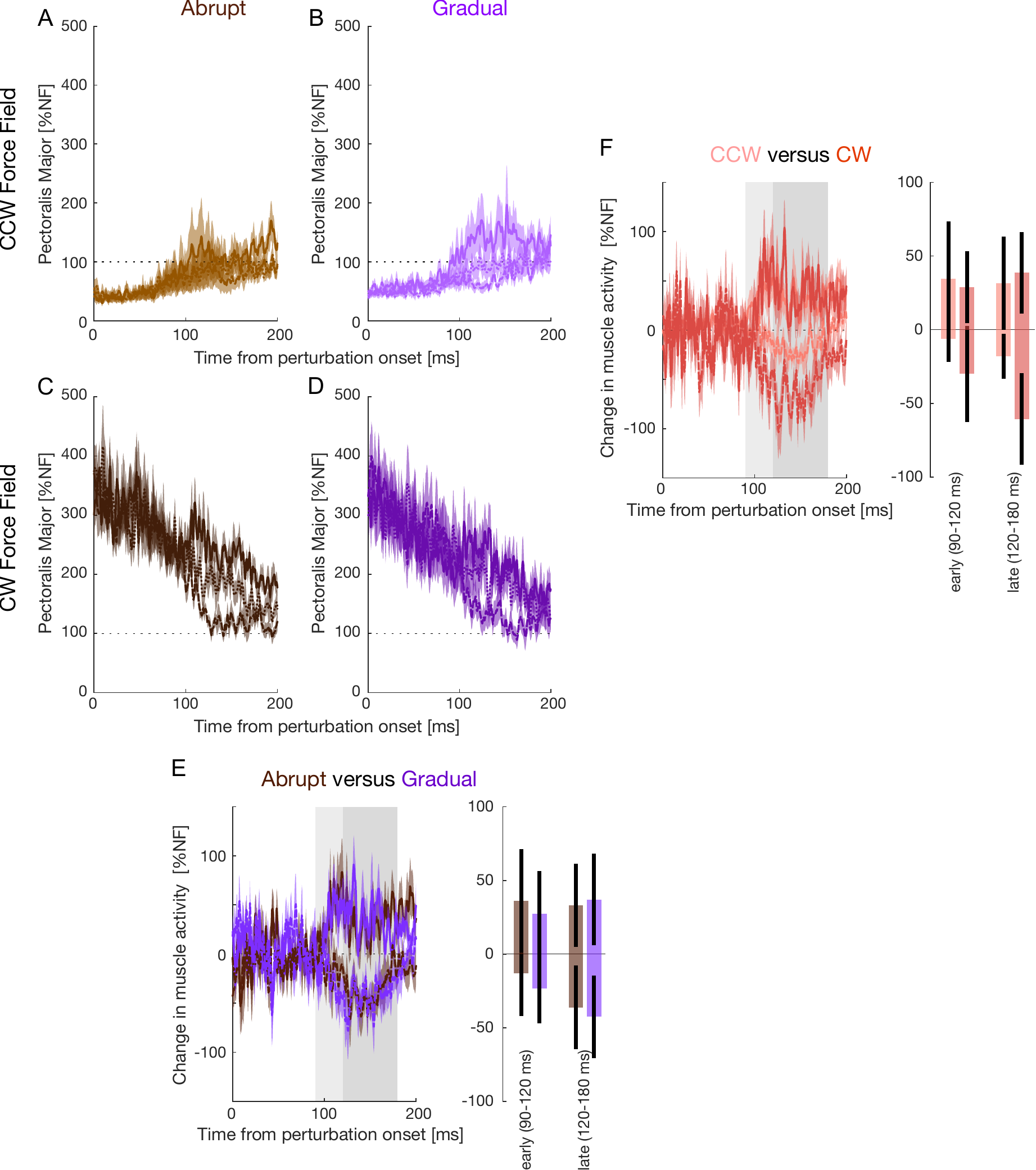}
\caption{Visuomotor feedback responses in the pectoralis major muscle after force field adaptation measured on probe trials. A: Pectoralis major activity to leftward (dashed lines), zero (dotted lines), and rightward (solid lines) visual perturbations after abrupt adaptation to the CCW force field. Activity is scaled according to the level of muscle activity in the null field (mean between -50 and +50 ms prior to the perturbation time) which is represented by the dotted black line. Shaded region indicates the s.e.m. B: Muscle activity after gradual adaptation to the CCW force field. C: Muscle activity after abrupt adaptation to the CW force field. D: Muscle activity after gradual adaptation to the CW force field. E: Visuomotor responses (perturbation – zero perturbation) averaged across the CCW and CW force fields for the abrupt (brown) and gradual (purple) conditions. Rightward perturbations (solid lines) produce an excitatory response whereas leftward perturbations (dashed lines) inhibit the muscle activity. Light grey and dark grey bars indicate the early (90-120 ms after perturbation onset) and late (120-180 ms) visuomotor response time windows. Bar plot quantify the responses over the early and late windows. Error bars indicate 95\% confidence intervals. F: Visuomotor responses averaged across abrupt and gradual conditions to examine differences between the CCW (pink) and CW (dark red) force fields}
\end{figure}

\begin{figure}
\centering
\includegraphics[width=13cm]{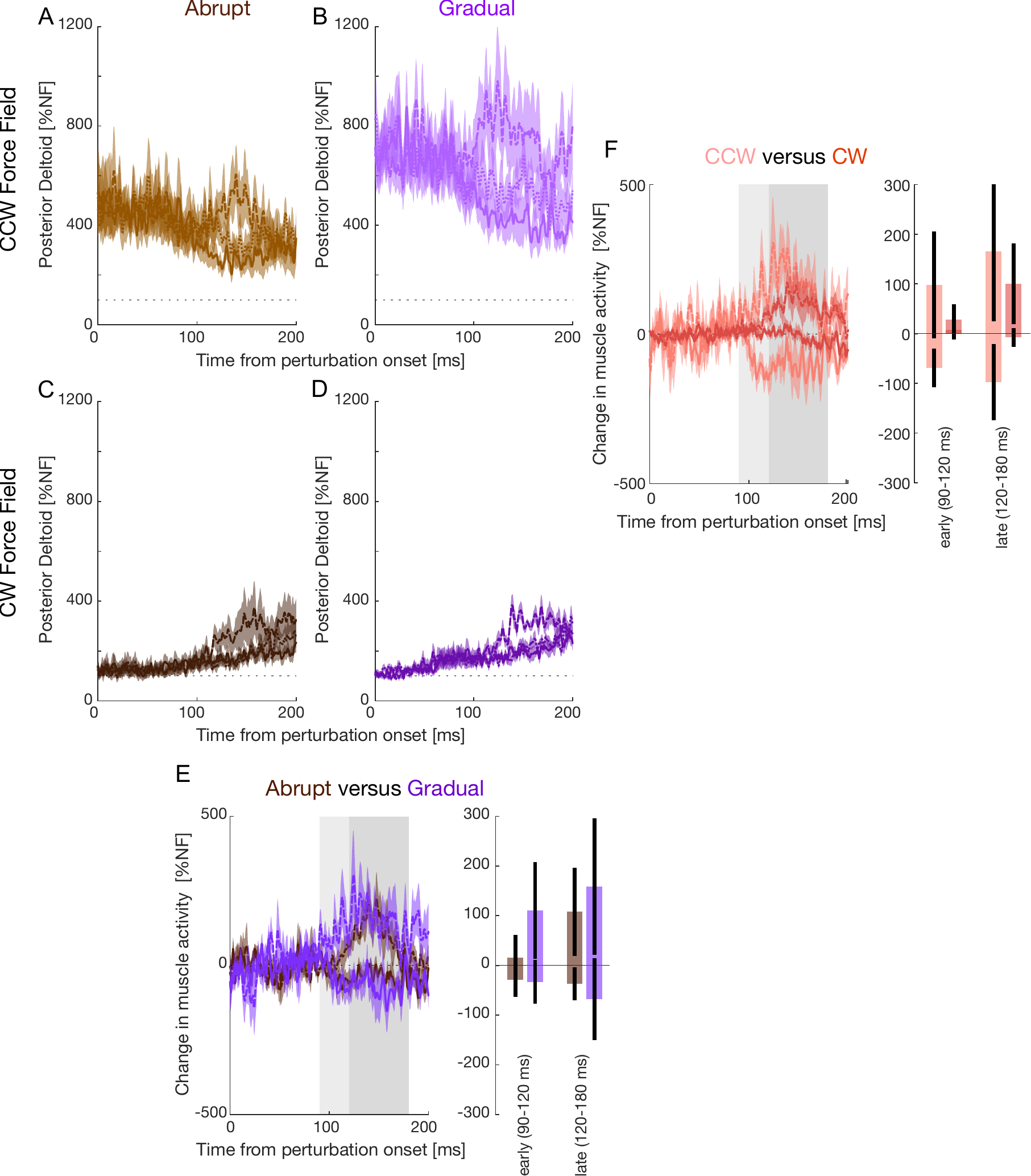}
\caption{Visuomotor feedback responses in the posterior deltoid muscle after force field adaptation measured on probe trials. Responses plotted as in Fig. 8 to leftward (dashed lines), zero (dotted lines), and rightward (solid lines) visual perturbations. A: Posterior deltoid activity to leftward, zero, and rightward visual perturbations after abrupt adaptation to the CCW force field. B: Gradual adaptation to the CCW force field. C: Abrupt adaptation to the CW force field. D: Gradual adaptation to the CW force field. E: Visuomotor responses (perturbation – zero perturbation) averaged across the CCW and CW force fields for the abrupt (brown) and gradual (purple) conditions. Bar plot shows responses over the early and late windows, with error bars indicating 95\% confidence intervals. F: Visuomotor responses averaged across abrupt and gradual conditions to examine differences between the CCW (pink) and CW (dark red) force fields.}
\end{figure}

We then examined the muscle responses to the visual perturbations in the pectoralis major (Fig. 8) and posterior deltoid (Fig. 9); the primary muscles to provide corrective responses to lateral perturbations for these movements. The muscle responses to the visual perturbation are shown separately for CCW and CW fields (Fig. 8A-D) as the background load and muscle activity are different across force fields. Visual perturbations produce clear changes in muscular activity. Averaging across the two force fields shows similar responses in the pectoralis major in the abrupt and gradual conditions (Fig. 8E), with no differences across either the early or late visuomotor response windows (error bars represent 95\% confidence intervals). Averaging across the abrupt and gradual conditions directly compares the muscular responses in the CCW versus the CW force fields (Fig. 8E). Here there are clear differences across the whole temporal response, with larger excitation and inhibition responses in the CW force field, although the 95\% confidence intervals mostly overlap across both early and late intervals. Relatively similar responses were observed in the posterior deltoid (Fig. 9). When averaged across the force fields, we found little differences in the muscular responses between the abrupt and gradual conditions (Fig. 9E). However, when averaging across abrupt and gradual conditions, we now found slightly larger feedback responses in the CCW force field compared to the CW force field, although again the 95\% confidence intervals were very large (Fig. 9F). That is, in the two force fields there is some evidence for larger muscular responses (both excitation and inhibition) when the muscle was more active resisting the force field. However, we do not have sufficient data to statistically test this possible difference here.

\subsection{Comparison against adaptation with full visual information}

\begin{figure}
\centering
\includegraphics[width=10cm]{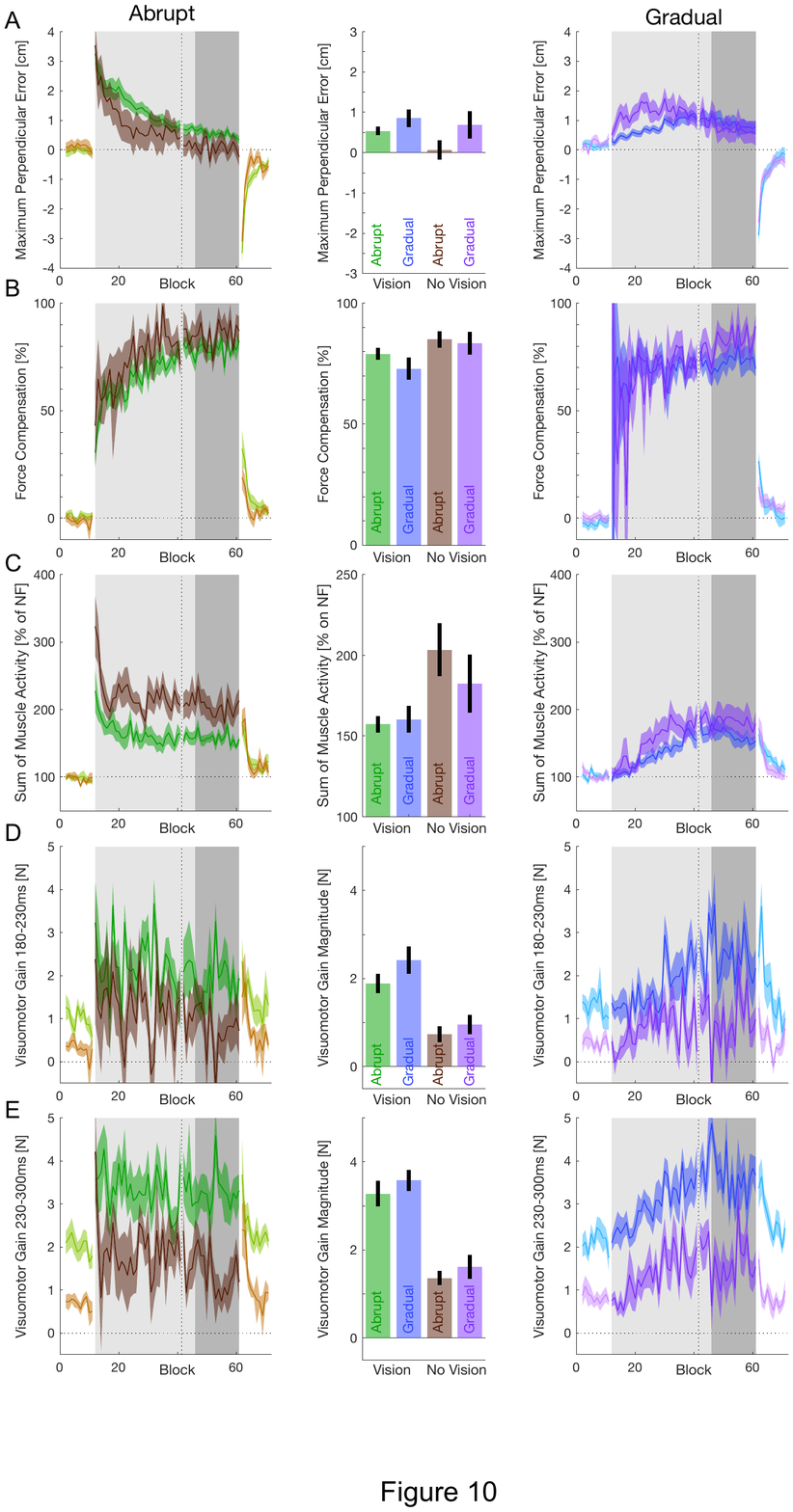}
\caption{Comparison of adaptation measures from the current experiment (brown and purple traces) with previous results from an experiment with full visual error feedback \cite{franklin_feedback_2021} (green and blue traces). A: Maximum perpendicular error during abrupt (left) and gradual (right) exposure to curl force fields. The bar plot illustrates final exposure levels during the final 15 blocks (dark shaded region). B: Force Compensation. C: Sum of Muscle Activity from 6 muscles, plotted relative to null field level. D: Early visuomotor feedback gain. E: Late visuomotor feedback gain. }
\end{figure}

Participants adapted to curl force fields with no lateral visual information regarding the error experienced during the reaching movement. In our previous work participants performed the same series of experiments but with full visual information \cite{franklin_feedback_2021}. Despite these differences, the overall pattern of adaptation and changes in visual feedback gains were similar. Here we compare the final measures to see whether there is any effect of limiting lateral visual error information on adaptation. For statistical comparison we ran frequentist and Bayesian ANOVAs on the data with main factors of visual feedback (2 levels: presence or absence), and exposure type (2 levels: abrupt or gradual exposure) in JASP. The data with the presence of visual feedback was taken from the previously published experimental work \cite{franklin_feedback_2021}.

The maximum perpendicular kinematic error during the experiments was similar with both the presence or absence of lateral visual feedback during the movements (Fig. 10A). However, in the abrupt condition we found that the MPE was slightly smaller in the absence of lateral visual feedback throughout the whole exposure period. In contrast, in the gradual condition the MPE was higher in the absence of lateral visual feedback in the initial exposure period although this disappeared by the end of the full exposure period. We compared the level of final adaptation across all conditions and found no significant main effects of visual feedback ($F_{1,44}=1.748, p=0.193$), exposure type ($F_{1,44}=3.699, p=0.061$), or interaction effects ($F_{1,44}=0.388, p=0.537$), suggesting that the final levels were similar across conditions. This was supported with the Bayesian ANOVA which found little or no support for any effect of visual feedback or exposure type (best fit model included only exposure type, $BF_{10}=1.258$) 

Although the increase in force compensation was similar across all conditions (Fig. 10B), we found that the final levels of force compensation were slightly higher with no lateral visual error feedback than the previous study with visual feedback ($F_{1,44}=4.402, p=0.042$), but no effect of exposure type ($F_{1,44}=0.974, p=0.329$) or interaction effect ($F_{1,44}=0.317, p=0.576$). The Bayesian ANOVA provided little support for any difference across conditions, with the best-fit model including only the visual feedback ($BF_{10}=1.707$).

In order to compare a single measure across the conditions we calculated the total muscle activation (sum of six muscle activities) during the experiments. Although the pattern of adaptation is similar regardless of the presence or absence of visual lateral error in both the abrupt conditions and in the gradual conditions, the final level of muscle activation at the end of the exposure period appears different (Fig. 10C). We compared overall muscle activation and found a main effect of visual feedback ($F_{1,44}=6.848, p=0.012$) but no effect of exposure type ($F_{1,44}=0.469, p=0.497$) or interaction ($F_{1,44}=0.843, p=0.364$). This was supported with the Bayesian ANOVA which found the best fit model to only include the main effect of visual feedback ($BF_{10}=4.397$). Specifically, we find that there was larger overall muscle activation when the visual lateral error information was not presented to the participants. 

The most prominent difference between the experiments was found for the visuomotor feedback responses. Here we found clear differences in the magnitudes of the visuomotor feedback responses, with higher values when lateral visual error information was present in both early and late time windows across the entire experiment as well as in the final exposure period (Fig. 10 D,E). Comparing the initial exposure levels (first 5 blocks) across conditions shows a main effect of both visual feedback ($F_{1,44}=8.142, p=0.007$) and exposure type ($F_{1,44}=14.845, p<0.0.001$), but no interaction effect ($F_{1,44}=0.173, p=0.679$) for the early window. A Bayesian ANOVA confirmed this finding with strong evidence for a model including both visual feedback and exposure type ($BF_{10}=277.741$). Similarly, the late window showed a main effect of visual feedback ($F_{1,44}=17.909, p<0.001$) and exposure type ($F_{1,44}=16.103, p<0.001$) but not an interaction ($F_{1,44}=0.049, p=0.827$) for the early window. A Bayesian ANOVA confirmed this finding with strong evidence for a model including visual feedback and exposure type ($BF_{10}=5638.551$). However, statistical comparison of the final exposure levels found a main effect of visual feedback ($F_{1,44}=30.347, p<0.001$), but no effect of exposure type ($F_{1,44}=2.476, p=0.123$) or interaction ($F_{1,44}=0.440, p=0.511$) for the early window. A Bayesian ANOVA confirmed this finding with strong evidence for a model including only visual feedback ($BF_{10}=7311.1$). Similarly, the late window showed a main effect of visual feedback ($F_{1,44}=61.810, p<0.001$), but no effect of exposure type ($F_{1,44}=1.270, p=0.266$) or interaction ($F_{1,44}=0.008, p=0.929$) for the early window. A Bayesian ANOVA confirmed this finding with strong evidence for a model including only visual feedback ($BF_{10}=1.837x10^{7}$). Therefore, as predicted by our theory regarding internal model uncertainty driving feedback modulation, we find that while early exposure produces large increase in visuomotor feedback responses in the abrupt condition but not in the gradual condition, whereas by the end of adaptation but the abrupt and gradual conditions have similar levels of modulation. The presence or absence of visual feedback simply scales this modulation.

Overall, there is a clear evidence that visuomotor feedback responses are down-regulated when the lateral visual error information is not presented to participants, whereas the overall muscle activity was increased in these conditions.

\section{Discussion}

The goal of this study was to examine whether online visual error information is important in controlling the changes in visuomotor feedback gains during dynamic adaptation. In particular, we recently claimed that the initial reactive visuomotor feedback gains increase due to internal model uncertainty, explaining the rapid increase in visuomotor feedback gains during an abrupt introduction of dynamics. However, it may have been simply the presence or absence of large visual errors, rather than the internal model uncertainty. Here we test this directly by having participants perform reaching movements where the environmental dynamics are either introduced abruptly or gradually in a design identical to our previous study \cite{franklin_feedback_2021} with the exception that participants received no online lateral visual error information. In the initial exposure to the force fields, participants in the abrupt condition experienced large kinematic errors, large muscle co-contraction, increased visuomotor feedback gains, and a rapid increase in the force compensation, whereas in the gradual condition there were little or no kinematic errors, little co-contraction, and only small increases in the visuomotor feedback gains. At the end of adaptation, similar levels were found across all measures in the abrupt and gradual conditions. Overall, we found similar general changes in all measures with abrupt and gradual adaptation to our previous study. However, we when contrasted the results of the two studies we found two major differences. When visual error information was removed, we found increases in muscle co-contraction and a reduction in the visuomotor feedback gains. That is, it appeared that in the absence of visual error signals, participants relied more on co-contraction and less on visuomotor reflexes. Overall, our work further supports the idea that reactive feedback gains may reflect internal model uncertainty and are not driven purely by visual error signals.

Many studies that have examined the role of online visual feedback on the adaptation process of motor control. It has been clearly established that online visual feedback is not required for adaptation to stable or unstable novel dynamics \cite{dizio_congenitally_2000, scheidt_interaction_2005, franklin_visual_2007, tong_kinematics_2002}. These have shown that the rate and extent of adaptation is similar both with and without online visual feedback, although endpoint visual feedback is important for learning the appropriate direction of the movement \cite{scheidt_interaction_2005}. These have suggested that the presence of proprioceptive signals can be used to drive rapid adaptation to novel dynamics. On the other hand, it has also been shown that even in the absence of proprioceptive signals, people are able to adapt to these novel dynamics. This has been shown both for deafferented participants \cite{sarlegna_force-field_2010, lefumat_generalization_2016, yousif2015proprioception} and under conditions where only visual signals are provided \cite{melendez-calderon_force_2011, zhou2022motion}. Here we examined the adaptation to abrupt and gradual changes in environmental dynamics in the absence of online lateral error information, but in the presence of final endpoint error. We found no major differences in the adaptation in terms of kinematic error or predictive force compensation to our previous study in which full visual error information was provided. The presence of the final endpoint visual location provided at the end of each movement was sufficient to produce adaptation of the direction of reach towards the target. Despite the fact that we found no major differences in adaptation with and without visual error, different error signals could potentially drive the adaptation in these two scenarios. Specifically, clamping the visual error information must induce a different sensory prediction error (at least for visual feedback) throughout the movement. As sensory prediction error has been highlighted as a major driver of adaptation \cite{shadmehr_error_2010, tseng_sensory_2007, izawa_learning_2011}, there is a question about how this might influence adaptation. However, given the similarities in adaptation across visual conditions, it is likely that the relative weighting of visual to proprioceptive feedback in the clamped condition is reduced, either in the estimation of a sensory prediction error, or in the effect of different sensory prediction errors driving learning. Regardless, this work further supports previous findings that online visual feedback of errors is not necessary for adaptation to novel dynamics.

The importance of the visual feedback on motor learning is well studied. Although the visual feedback is not essential for motor adaptation \cite{franklin_visual_2007, tong_kinematics_2002, dizio_congenitally_2000}, it appears to be responsible for learning the direction of the movement and path planning \cite{scheidt_interaction_2005}. Moreover, when visual cues are congruent with the underlying dynamics, there is evidence for faster adaptation \cite{franklin2023congruent} and better online control \cite{vcesonis2019controller, franklin_influence_2018}. The fact that adaptation to visuomotor rotations can occur successfully even in the absence of proprioceptive feedback \cite{bernier_updating_2006, tsay2023implicit} shows that the visual signal enables a remapping of the movement direction plan. However the type of visual feedback affects the adaptation that occurs, with differences in the adaptation produced by continuous visual feedback to only terminal visual feedback \cite{heuer2008constraints}. Indeed it has been suggested that the presence of terminal visual feedback is important to generate reward prediction errors and drive implicit adaptation \cite{leow2018task, ikegami2021hierarchical}. In addition to driving adaptation, visual feedback also provides important information for dynamical control, specifically to select different internal models of objects \cite{gordon_memory_1993}, or to select different internal models for different dynamic force fields \cite{howard_effect_2013, hirashima_distinct_2012}. 

We claimed that internal model uncertainty drives the changes in reactive visuomotor feedback gains, whereas the predictive feedback gains are learned slowly as part of the adaptation process \cite{franklin_feedback_2021}. The interpretation that internal model uncertainty drives reactive feedback gains is supported by several lines of evidence. First, that sudden changes in the dynamics (either adding to or removing from the environment) which produce large errors drive these increased feedback gains \cite{franklin_visuomotor_2012, franklin_feedback_2021, maurus2023nervous}. Second, that these increased feedback gains are gradually reduced as learning occurs which would parallel the reduction in internal model uncertainty. These changes in feedback gains are similar to those in grip forces during adaptation, in which the grip force is most sensitive to variability in the magnitude of the force fields rather than the strength of the force fields themselves \cite{hadjiosif_flexible_2015}. However, there was potentially another explanation for our data, that it was the large visual errors that were driving rapid increases in the visuomotor gains when the dynamics changed abruptly. Here we tested this directly by removing the lateral visual information of kinematic errors using visual clamp trials. Despite an overall reduction in the visuomotor feedback gains across all experiments, the pattern of visuomotor feedback gains was the same as with visual feedback. Despite the fact that visual information could not be used for adaptation to the force fields in the current study, we still found increases in the visuomotor feedback gains that paralleled those in our previous study \cite{franklin_feedback_2021}. This further supports our hypothesis that internal model uncertainty drives the rapid changes in the reactive feedback gains.

When visual feedback of the lateral errors was removed in this study, we found a consistent reduction in the visuomotor feedback responses across all phases of the experiment relative to when it was present (Fig. 10 D,E). This suppression of the visuomotor feedback responses was seen even at the beginning of the pre-exposure phase. Prior to this phase, every participant performed a familiarization phase where they were first exposed to the presence or absence of visual error. As has been shown previously, the presence of task-relevant visual errors produces increases in visuomotor feedback responses whereas the absence of such errors (or the presence of task-irrelevant visual errors) decreases these feedback gains \cite{franklin_specificity_2008, franklin_fractionation_2014}. When visual feedback is clamped, as in the current study,the sensorimotor control system likely down-regulates the visuomotor feedback gains, as visual feedback of lateral error is determined to be irrelevant to the task. However this does not mean that participants do not use visual feedback, as they require this to stop within the final target position without overshooting. Instead, as we have previously demonstrated \cite{franklin_specificity_2008, franklin_fractionation_2014}, the sensorimotor control system modulates the overall gains of the visuomotor feedback system according to the relevance of such visual error for task success. 

As discussed in our previous work \cite{franklin_feedback_2021}, here we claim that it is internal model uncertainty rather than internal model inaccuracy that drives the changes in reactive feedback gains. While we do not directly measure internal model uncertainty, we argue that this drives the feedback gain modulation for the following reasons. Any movement in which the dynamics are suddenly changed, will have large internal model inaccuracy and kinematic errors, but initially low internal model uncertainty until late in the movement. Here we would predict initially low reactive feedback gains on this movement, but high feedback gains on subsequent movements. Alternatively we predict higher feedback gains if the internal model uncertainty is high, even if the internal model accuracy is perfect and there are no errors on a specific trial (as might occur when the external dynamics vary from trial to trial).

In contrast to the reactive component, the predictive component is adapted to the environment. We propose that part of adaption to novel environments or tasks is the gradual tuning of the feedback gains to the specific task or action being performed \cite{franklin_feedback_2021}. After adaptation, feedback gains have been shown to tune appropriately to both novel dynamics \cite{cluff_rapid_2013, franklin_visuomotor_2012, franklin_rapid_2017, franklin_endpoint_2007}, and novel visual environments \cite{franklin_specificity_2008, franklin_fractionation_2014, hayashi2016visuomotor}. In our current study we found similar changes of the visuomotor feedback gains during adaptation to the force fields for both the abrupt and gradual conditions despite the absence of the visual error indicating the force field. This suggests that proprioceptive information is sufficient to drive these learned changes in the visuomotor gains, tuning them to the dynamics. As recent work has shown that vestibular inputs can modulate visuomotor feedback gains \cite{oostwoud2019vestibular}, it is likely that the sensorimotor system integrates multi-sensory information \cite{debats2021visuo, ernst_humans_2002, smeets_sensory_2006} to drive changes in feedback gains. Although the overall magnitude of the visuomotor feedback gains was reduced overall when no visual signals of the lateral error was provided in the current experiment, the increase relative to the null field was similar with or without visual feedback (Fig. 10 D,E). This increase, despite the fact that visual errors about the force field were never presented to the participants suggests that the gradual learning of the predictive feedback gains may rely on a mixture of sensory signals indicating the changes in dynamics.  

Most studies examining the adaptation of feedback gains and the learning of feed-forward control have suggested a parallel adaptation process with shared internal models \cite{cluff_rapid_2013, coltman_time_2020, wagner_shared_2008, ahmadi-pajouh_preparing_2012, maeda_feedforward_2018, franklin_endpoint_2007}. However, it has also been suggested that feedforward and feedback control are learned separately in a mirror adaptation task \cite{kasuga_learning_2015}. In this study, participants produced stronger feedforward adaptation when feedback corrections were not allowed (removal of online visual feedback). One possibility is that the feedforward adaptation in this task (assessed using initial movement direction) may have been primarily due to explicit changes in strategy, as often found in visuomotor adaptation experiments \cite{hegele_implicit_2010, bond2015flexible}. In contrast, most adaptation to force fields appears to be implicit in nature \cite{schween2020assessing, forano2021direct}. One could therefore argue that most studies support shared internal model adaptation of both the implicit feedforward and feedback controllers. We propose that this describes the adaptation of the predictive feedback controller which likely shares structures and internal models with the implicit feedforward controller. In contrast, we expect that reactive feedback gains may be independent of this process. A recent study \cite{cesonis2022contextual} has shown context dependent switching of feedback controllers for different tasks similar to the contextual switching of feedforward controllers for dual adaptation \cite{howard_gone_2012, howard_value_2015, forano2020timescales, heald2021contextual, oh_minimizing_2019, howard2020asymmetry}. Such findings closer connect the adaptation of feedforward and feedback controllers, although clear evidence of the development and switching of these independent feedback controllers during learning has yet to be demonstrated. While our current work cannot specifically test these ideas, the gradual adaptation of the predictive feedback gains in the gradual force field parallels the increased force compensation throughout the experiments, similar to what has been seen in other studies \cite{franklin_feedback_2021, cluff_rapid_2013}. This supports the idea that predictive feedback gains and feedforward commands adapt on a similar time scale and may indeed share internal models as previously proposed \cite{wagner_shared_2008, ahmadi-pajouh_preparing_2012, maeda_feedforward_2018}.

\begin{figure}[t]
\centering
\includegraphics[width=10cm]{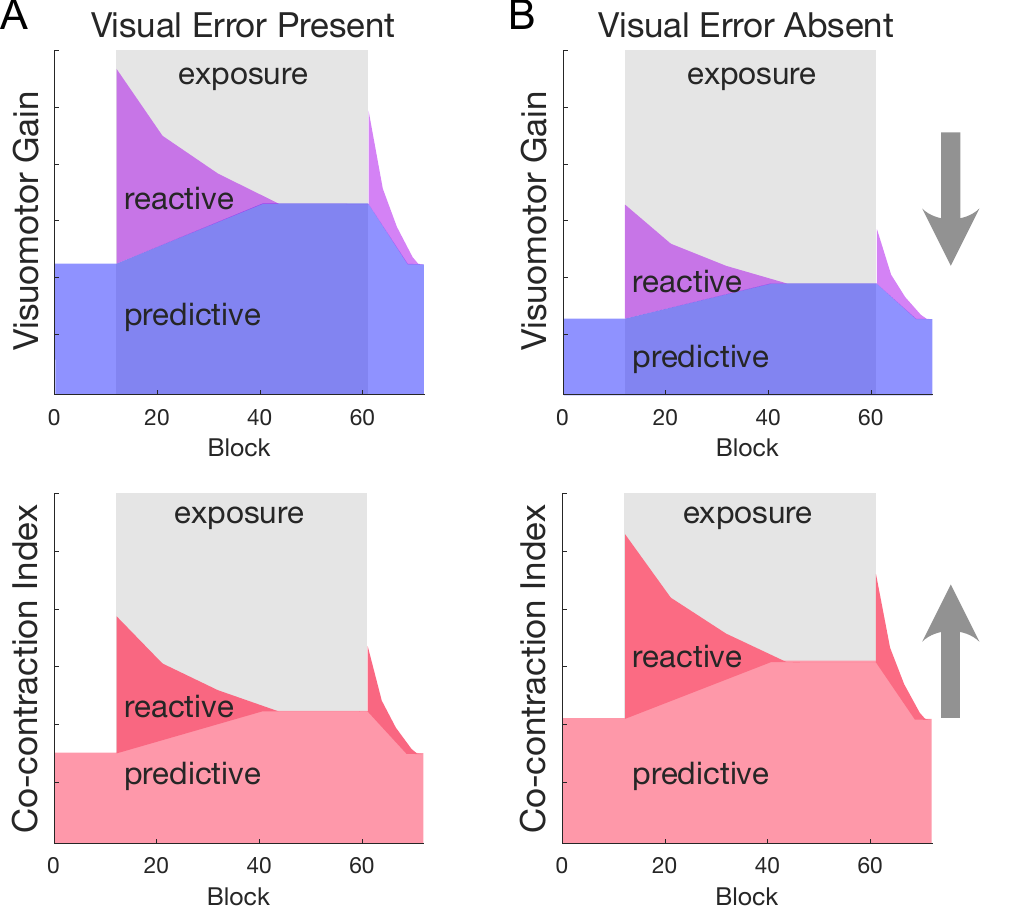}
\caption{Schematic of changes in visuomotor feedback gains and co-contraction  during adaptation with and without visual error information. A: Presence of visual error information. The pattern of feedback gain modulation (upper plot) and co-contraction (lower plot) during adaptation and de-adaptation to changes in environmental dynamics with visual error information. We propose that the total feedback gain is comprised of the reactive (purple) and predictive (blue) changes in feedback gain during adaptation \cite{franklin_feedback_2021}. Changes in co-contraction exhibit a similar pattern of modulation with changing dynamics that we propose is also comprised of a reactive response to sudden errors (dark red), and a predictive response that is learned and tuned to the dynamics (light red). B: Absence of visual error information. Despite the removal of visual error information we find a similar pattern of modulation of visuomotor feedback gains with the changing dynamics. However the overall level is lower indicating a general suppression of the visuomotor feedback gains. In contrast the co-contraction increases to compensate for the loss of the visual error information. This shows a tradeoff between co-contraction and feedback gains depending on the availability of the sensory information.}
\end{figure}

One major difference between the current results without lateral visual error and our previous study \cite{franklin_feedback_2021} that included full visual feedback was the increased muscle co-activation throughout the adaptation process. This increased muscle activation was still present even at the end of learning, and occurred regardless of the abrupt or gradual introduction of the force field (Fig 10 C). One clear finding is that the increased visuomotor feedback gains are therefore also independent of the background co-contraction muscle activity, as instead we find increased co-contraction when the visuomotor feedback gains are reduced. Co-contraction, which occurs rapidly upon the onset of novel dynamics \cite{osu_short-_2002, franklin_adaptation_2003, thoroughman_electromyographic_1999}, increases limb stiffness \cite{franklin_functional_2003, franklin_endpoint_2007} to reduce the perturbing effects of the force fields. Increased co-contraction of muscles has been shown to speed adaptation to novel dynamics \cite{heald_increasing_2018}, possibly by reducing the difference between the experienced dynamics and those that are required for final adaptation (e.g. \cite{gonzalez_castro_binding_2011}). It is likely that visuomotor feedback responses assist in a similar manner, correcting errors in the movements such that the adaptive controller can learn the appropriate feedforward compensation to the novel dynamics. Thus higher feedback gains could also drive faster adaptation to novel dynamics. This means that when such visual error information is not present, increased co-contraction is used to limit the disturbing effects of the novel dynamics. This would explain the tradeoff we find between co-contraction and visuomotor feedback gains in the two studies examining adaptation with an without visual feedback of errors.     

Overall, when we examine the feedback gain modulation upon introduction of novel dynamics, we find a consistent pattern of modulation: an initial rapid increase associated with large kinematic errors, followed by a gradual reduction to a new baseline level during the exposure (Fig 11). We have suggested in our previous paper that this typical pattern is composed by two complementary processes which we termed as reactive and predictive feedback gains \cite{franklin_rapid_2017, franklin_feedback_2021}. We have predicted \cite{franklin_feedback_2021} that each of these two components has different properties; reactive feedback gains are likely to be broader in terms of temporal timing whereas predictive feedback gains may be more likely to generalize spatially to nearby movements similar to the generalization of predictive force \cite{shadmehr_adaptive_1994, berniker_motor_2014, malfait_transfer_2002, leib2021error, orschiedt2023learning, donchin_quantifying_2003, thoroughman_learning_2000}. We propose that the reactive feedback gains correspond to a robust control policy that attempts to provide stability despite internal model inaccuracies, similar to reactive increases in co-contraction, and that the predictive feedback gains could reflect a shifting of the control policy to an energy efficient mode as the internal model accuracy is improved, paralleling equivalent changes in the feedforward controller. In this study, we demonstrated the the absence of visual error information produces similar patterns of adaptation of feedforward and feedback controllers, but with increased co-contraction to compensate for the reduced visuomotor feedback responses.

%\section*{ACKNOWLEDGEMENTS}

\section*{CONFLICT OF INTEREST}
The authors declare no competing financial interests.

\section*{DATA AVAILABILITY}
Data and code for analysis is available at: 10.6084/m9.figshare.23271833

\printendnotes

% Submissions are not required to reflect the precise reference formatting of the journal (use of italics, bold etc.), however it is important that all key elements of each reference are included.
\bibliography{main}

%\begin{biography}[example-image-1x1]{A.~One}
%Please check with the journal's author guidelines whether author biographies are required. They are usually only included for review-type articles, and typically require photos and brief biographies (up to 75 words) for each author.
%\bigskip
%\bigskip
%\end{biography}

%\graphicalabstract{example-image-1x1}{Please check the journal's author guildines for whether a graphical abstract, key points, new findings, or other items are required for display in the Table of Contents.}

\end{document}